\documentclass[prd, floatfix,
 reprint,
superscriptaddress,
groupedaddress,
unsortedaddress,
nofootinbib,
 amsmath,amssymb,
 aps,
]{revtex4-2}
\usepackage{amsmath}
\usepackage{amssymb}

\usepackage{graphicx}
\usepackage[dvipsnames]{xcolor}
\definecolor{light gray}{gray}{0.8}
\usepackage{pgfplots}
\usepackage{tikz}
\usetikzlibrary{decorations.markings} 

\usepackage{placeins}
\usepackage{dcolumn}
\usepackage{bm}
\definecolor{prdlink}{RGB}{0,0,128}\usepackage[colorlinks=true,linkcolor=prdlink,citecolor=prdlink,urlcolor=prdlink]{hyperref}
\usepackage{orcidlink}

\pgfplotsset{compat=1.18}
\begin{document}

\preprint{APS/123-QED}

\title{Deflection of light and time delay in a hyperbolic Einstein-Straus–de Sitter solution}

\author{Mourad Guenouche\,\orcidlink{0000-0002-1211-4940}}
  \email{guenouche\_mourad@umc.edu.dz}\email{guenouche\_mourad@univ-khenchela.dz}
  
\affiliation{Laboratoire de Physique Théorique, \href{https://ror.org/017wv6808}{Université Frères Mentouri-Constantine 1}, BP 325 route de Ain El Bey, 25017 Constantine, Algeria \\
and Département des Sciences de la Matière, \href{https://ror.org/02yyskm09}{Université Abbès Laghrour-Khenchela}, BP 1252, El Houria, Route de Constantine, 40004 Khenchela, Algeria}

\date{\today}

\begin{abstract}
We analyze strong gravitational lensing by a spherically symmetric mass distribution within the Einstein-Straus–de Sitter framework in a spatially open universe with negative curvature ($k = -1$). Applying the theory to the lensed quasar SDSS J1004+4112, we identify a critical threshold for the current scale factor $a_0$ of approximately $2.6\times10^{27}\,\text{m}$, below which the effects of negative spatial curvature on lensing observables become significant, corresponding to a current curvature density of $|\Omega_{k0}|\gtrsim0.0025$. In particular, for $\Omega_{k0}=-0.15$, the light bending increases slightly by $\sim 1\%$, while the time delay exhibits a more substantial increase of $\sim 10\%$. Beyond this threshold, however, the lensing observables are found to be insensitive to the current scale factor and converge to those characteristic of a spatially flat universe. Importantly, our results indicate that, even in scenarios where spatial curvature would otherwise enhance lensing observables, the effect of the cosmological constant does remain present and acts to reduce light bending, corroborating the claim of Rindler and Ishak.
\end{abstract}

\maketitle
\section{Introduction}
Gravitational lensing provides a powerful probe of cosmic structures, but its interpretation relies critically on assumptions about the Universe's geometry. While current cosmological observations suggest that the spatial curvature of the Universe is extremely close to zero \cite{Planck2020,DiValentino2021}, the possibility of a small but nonzero curvature remains open. Most lensing studies, however, assume perfect spatial flatness, allowing for the use of Euclidean geometry to estimate light deflection and time delays. Such simplifications, though convenient, break down in curved spacetimes, where photon paths deviate from straight lines even outside local inhomogeneities, potentially biasing predictions. Studying spacetimes with negative spatial curvature is therefore essential for building a comprehensive picture of all possible Friedmann–Lemaître–Robertson–Walker (FLRW) cosmologies. Such deviations from flatness are particularly relevant given that generic early-universe scenarios and approaches in quantum cosmology often predict slight negative curvatures as a natural outcome \cite{Linde1982,Linde1986,Freivogel2006}. Moreover, inflationary and other theoretical models do not guarantee exact flatness \cite{Guth1981,Planck2020}, and so small levels of a hyperbolic geometry are plausible. An open universe ($k = -1$) would not only alter the global geometry but also influence the propagation of light rays over cosmological distances, directly affecting gravitational lensing observables. Compared to flat or positively curved cases, negative curvature tends to increase comoving distances and enhance the divergence of geodesics, leading to subtle but systematic differences in predicted deflection angles and time delays. Quantifying these deviations provides a pathway to test the robustness of standard cosmological models and to explore whether future observations could place tighter constraints on $\Omega_{k0}$.

The Einstein-Straus–de Sitter (ESdS) model, with a cosmological constant $\Lambda$, is a foundational example of a Swiss-cheese cosmology \cite{ES1,ES2,Balbinot}, offering a self-consistent framework for studying realistic astrophysical phenomena. It elegantly reconciles cosmic expansion on large scales with local static gravitational fields by embedding a static Schwarzschild–de Sitter (SdS or Kottler) metric, which describes local systems such as galaxies or clusters, within a dynamic FLRW universe.

This work extends the analysis of strong gravitational lensing within the ESdS model to specifically examine the case of negative spatial curvature $k=-1$, which, despite being included in general treatments \cite{Guen3}, remains largely underexplored in detail in the literature. Our calculations are grounded in the core principles of general relativity and are based on direct integration of null geodesic equations. This approach accommodates arbitrary spatial curvature, without recourse to flat space approximations to compute light bending and time delay.

Motivated by ongoing debate regarding the influence of the cosmological constant $\Lambda$ on light deflection \cite{Rind,Ishak,Schu4,Schu6,Sereno,Khrip,Park,Simpson,Kant1,Schu5,Chen,Kant2,Arakida,Kant3,Sultana1,Heydari,Hu,Sultana2,Sultana3}, we aim to reexamine this issue in the context of a negatively curved ESdS model. While recent studies suggest that the effect is negligible, it remains an open question whether negative spatial curvature could modify or enhance this outcome. For instance, the authors of Ref~\cite{Arakida,Wang} assert that the contribution of $\Lambda$ is absorbed into the impact parameter or angular diameter distance, making its direct effect minuscule. In Ref.~\cite{Hu}, the authors find no extra measurable dependence of gravitational lensing observables on the cosmological constant beyond what is already present in the standard lensing equations within the Swiss-cheese cosmological model.

The lensed quasar SDSS J1004+4112 is selected as a test case to quantify the influence of the spatial curvature and the cosmological constant on gravitational lensing observables. We assess these effects through systematic variations in the current cosmic scale factor $a_0$. For direct comparison to prior analyses within flat ($k=0$) and positively curved ($k=+1$) ESdS models, presented in Refs.~\cite{Schu1,Guen1,Guen2}, we adopt the same cosmological parameters as employed therein.

The paper is organized as follows. Section~\ref{sec2} establishes the geometric framework of the hyperbolic ESdS model by presenting the junction conditions required to match the static Schwarzschild–de Sitter vacuole to an expanding FLRW universe with negative spatial curvature. In Sec.~\ref{sec3}, the detailed methodology for computing light deflection and time delay is developed, including direct integration of null geodesic equations across the various spacetime regions leading to relevant analytical expressions. Section~\ref{sec4} applies this formalism to the lensed quasar SDSS J1004+4112, systematically exploring how the lensing observables depend on the cosmic scale factor (or equivalently the curvature density) and the cosmological constant, and comparing the results with those obtained in flat and closed ESdS models.\footnote{All numerical evaluations are performed using Wolfram \textit{Mathematica} 11.} Finally, Sec.~\ref{sec5} summarizes the main findings and discusses their implications for cosmological lensing, as well as possible directions for future work. 

\section{\label{sec2}Junction conditions for hyperbolic Einstein-Straus--de Sitter metric}
In Schwarzschild coordinates $(T,r,\theta,\varphi)$, the static SdS metric
\begin{equation}
\text{d}s_{\text{SdS}}^{2}=B(r)\text{d}T^{2}-B(r)^{-1}\text{d}r^{2}-r^{2}
\left( \text{d}\theta ^{2}+\sin ^{2}\theta \text{d}\varphi ^{2}\right),
\label{SdS}
\end{equation}
with
\begin{equation}
B(r)=1-\frac{2GM}{r}-\frac{\Lambda }{3}r^{2},
\end{equation}
applies within a vacuole (Schücking sphere) of radius $r_{\text{Sch\"{u}}}(T)$ centered around a spherical mass distribution (the lens) $M $, $r\leq r_{\text{Sch\"{u}}}$. In the Friedmann coordinates $(t,r,\theta,\varphi)$, the dynamic FLRW metric, with a negative curvature constant $(k=-1)$,
\begin{equation}
\text{d}s_{\text{FLRW}}^{2}=\text{d}t^{2}-a(t)^{2}\left[\text{d}\chi ^{2}+\sinh
^{2}\chi \left( \text{d}\theta ^{2}+\sin ^{2}\theta \text{d}\varphi ^{2}\right)\right],\label{Hyp_FLRW}
\end{equation}
describes the spacetime outside the vacuole of constant radius $\chi _\text{Sch\"{u}}$, $\chi \geq \chi _\text{Sch\"{u}}$, where the scale factor $a(t)$ evolves according to the Friedmann equation,
\begin{equation}
\frac{\text{d}a}{\text{d}t}=f(a),\quad f(a)=\sqrt{\frac{A}{a}+\frac{\Lambda }{3}a^{2}-k},\label{FLRWeq}
\end{equation}
with $\rho$ the dust density and $A$ a constant coming from the energy conservation law for a nonrelativistic matter-dominated universe, $3A=8\pi G\rho a^3$. 

The two metrics are matched on the vacuole boundary under the matching condition
\begin{equation}
r_{\text{Sch\"{u}}}(T)=a(t)\sinh \chi _{\text{Sch\"{u}}},  \label{match}
\end{equation}
where $\sin\chi _{\text{Sch\"{u}}}=\left( 2GM/A\right) ^{1/3}$, using the fact that $M$ is related to $\rho$ by $3M=4\pi r_\text{Sch\"{u}}^{3}\rho$.

Following the same treatment outlined in~\cite{Schu1,Guen1,Guen2,Guen3}, the Jacobian of the transformation from the Schwarzschild to the Friedmann coordinates coordinates, $(T,r)\rightarrow (t,\chi )$, on the vacuole reads
\begin{equation}
\text{J}=\left( 
\begin{array}{cc}
\cosh \chi _{\text{Sch\"{u}}} & -\dfrac{f(a)\sinh\chi_{\text{Sch\"{u}}}}{B_{\text{Sch\"{u}}}} \\ 
-\dfrac{f(a)\sinh \chi _{\text{Sch\"{u}}}}{a} & \dfrac{\cosh \chi _{\text{Sch\"{u}}}}{aB_{\text{Sch\"{u}}}}
\end{array}
\right),\label{J}
\end{equation}
from which one can deduce a relation allowing to pass from the Schwarzschild time to the Friedmann time and vice versa,
\begin{equation}
\text{d}T=\cosh \chi _{\text{Sch\"{u}}}\frac{\text{d}t}{B_{\text{Sch\"{u}}
}(t)},\label{timematch}
\end{equation}
with $B_{\text{Sch\"{u}}}=B(r_{\text{Sch\"{u}}})$.
\section{\label{sec3}Light bending and time delay}
\begin{figure}[ht]
\centering
\begin{tikzpicture}

\fill[light gray] (0,0) circle [radius=1.5cm]; 
\fill[gray!62] (0,0) circle [radius=1.2cm]; 

\fill[black] (0,0) circle [radius=0.08cm] node[below, text=black]{L}; 
\draw[->, thin ] (-4.2,0) -- (4.2,0) node[above left] {$x$} ; 
\draw[dashed](0,0) -- (4.2,-0.5);
\draw[dashed, ->, thin](0,0) -- (0.75,1.29904) node[below right]{$\chi_\text{Schü}$} node[left]{$r_\text{Schü}$};

\fill[darkgray!85] (1.15911,0.310583) circle [radius=0.05cm]node[above, text=black]{$t_\text{SchüS}$};

\fill[darkgray!85] (1.03923,-0.6) circle [radius=0.05cm]node[below, text=black]{$t_\text{SchüS}^\prime$};

\fill[darkgray!85] (-1.42658,0.463525) circle [radius=0.05cm]node[above, text=black]{$t_\text{SchüE}$};

\fill[darkgray!85] (-1.35145,-0.650826) circle [radius=0.05cm]node[below, text=black]{$t_\text{SchüE}^\prime$};

\coordinate (O) at (4.2,-0.5);
\coordinate (A) at (1.15911,0.310583);
\coordinate (B) at (1.03923,-0.6);
\coordinate (C) at (-1.42658,0.463525);
\coordinate (D) at (-1.35145,-0.650826);
\coordinate (E) at (-4.2,0);
  
\draw[color=red, postaction={ decorate, 
decoration={ markings, mark=at position 0.55 with {\arrow{>}}}}] (O) to [bend left=5] (A);

\draw[color=red, postaction={ decorate, 
decoration={ markings, mark=at position 0.6 with {\arrow{>}}}}] (O) to [bend left=-5] (B);
  
\draw[color=red, postaction={ decorate, 
decoration={ markings, mark=at position 0.5 with {\arrow{>}}}}] (A) to [bend left=-10] (C);

\draw[color=red, postaction={ decorate, 
decoration={ markings, mark=at position 0.5 with {\arrow{>}}}}] (B) to [bend left=12] (D);

\draw[color=red, postaction={ decorate, 
decoration={ markings, mark=at position 0.6 with {\arrow{>}}}}] (C) to [bend left=5] (E);

\draw[color=red, postaction={ decorate, 
decoration={ markings, mark=at position 0.7 with {\arrow{>}}}}] (D) to [bend left=-5] (E);

\fill[red] (4.2,-0.5) circle [radius=0.08cm] node[below left, text=black]{S};

\fill[black] (-4.2,0) circle [radius=0.08cm] node[above right, text=black]{E};

\node at (1.7,-0.35) {$\chi_\text{LS}$};
\node at (-1.7,-0.2) {$\chi_\text{L}$};

\node at  (3.4,-0.15) {$-\varphi_\text{S}$};
\node at (-2.4,0.13) {$\alpha$};
\node at (-2.4,-0.15) {$\alpha^{\prime}$};

\draw[ thin, black] (3.05,0) arc (0:-20.5:1cm);
\draw[ thin, black] (-2.6,0) arc (0:11:1cm);
\draw[ thin, black] (-2.65,0) arc (0:-15.5:1cm);
\end{tikzpicture} 
\caption{Two photons emitted by a source S, bent inside the SdS vacuole by a lens L and finally received at Earth E. Outside the vacuole the trajectories diverge due to the pseudospherical geometry of negative curvature.}
\label{def}
\end{figure}
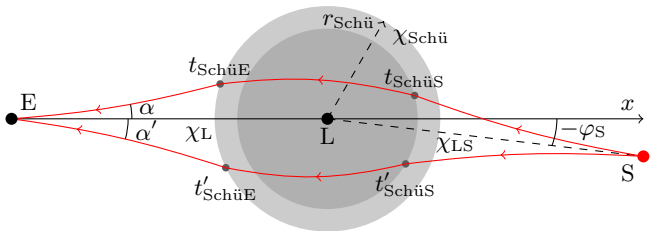

We consider two photons emitted from a source S at different times $t_\text{S}$ and $t_{\text{S}}^\prime$. These photons reach the boundary of the vacuole at $t _{\text{Sch\"{u}S}}$ and $t _{\text{Sch\"{u}S}}^\prime$, traverse the vacuole, and exit on the opposite side at $t _{\text{Sch\"{u}E}}$ and $t _{\text{Sch\"{u}E}}^\prime$. Both photons ultimately arrive simultaneously at Earth E at $t _{\text{E}}=t _{\text{E}}^\prime$, as depicted in Fig.~\ref{def}. This synchronization is motivated by the well-known observational conditions on Earth, which allow to express the time delay between the two photons as $\Delta t=t_{\text{S}}^\prime-t_{\text{S}}$. This implies a backward integration in time. Let $\alpha$ and $\alpha^\prime$ denote the angles that the photons make upon receipt on Earth, and $r_\text{P}$ and $r_{\text{P}}^\prime$ the minimum approach distances (perilens) where they get deflected by a lens L.

The Earth-lens and Earth-source comoving distances, denoted by $\chi_\text{L}=\chi(z_\text{L})$ and $\chi_\text{S}=\chi(z_\text{S})$ are calculated from the radial geodesic equation for given redshifts $z_\text{L}$ and $z_\text{S}$,
\begin{equation}
\chi (z)=\int_{\frac{a_0}{1+z}}^{a_{0}}\frac{da}{af(a)}.\label{comovdist}
\end{equation}
To obtain (\ref{comovdist}), I use the cosmological redshift relation $1+z=a_0/a$, where $a_0=a(t=0)$ is the current scale factor at Earth. The Earth-Source distance $\chi_\text{LS}$ is approximated by 
\begin{equation}
\chi_\text{LS}\simeq \chi_\text{S}-\chi_\text{L},\label{chiLS}
\end{equation}
due to the small source angular position $\varphi_\text{S}$.

\subsection{\label{subsecA}Integrating null geodesic equations between Earth and vacuole}
Although the FLRW background does not possess a single global center of spherical symmetry, it is homogeneous and isotropic, and thus spherically symmetric about every comoving observer regardless of the sign of curvature. The matched ESdS spacetime, constructed by embedding a spherically symmetric SdS vacuole into the FLRW universe, preserves this local spherical symmetry around the lens. This is ensured by the angular part of the metric, $\text{d}\theta^{2}+\sin^{2}\theta\text{d}\varphi^{2}$, which remains unchanged and continuous across the vacuole boundary, as shown explicitly in (\ref{SdS}) and (\ref{Hyp_FLRW}), and by the radial junction condition, Eq.~(5), which guarantees a smooth transition preserving spherical symmetry locally. Consequently, any photon geodesic can be confined, without loss of generality, to a single plane containing the lens, commonly chosen as the equatorial plane $\theta=\pi/2$. In this case, only the $t$ and $\varphi $ components of the FLRW null  geodesic equations are needed, which are
\begin{eqnarray}
&&\dot{t}\ddot{t}+a\dot{a}\left( \dot{\chi}^{2}+\sinh ^{2}\chi \dot{\varphi}^{2}\right) =0,  \label{geqFLRW1} \\
&&\frac{1}{2}\frac{\ddot{\varphi}}{\dot{\varphi}}+\frac{\dot{a}}{a}+\frac{%
\dot{\chi}}{\tanh \chi }=0,  \label{geqFLRW3}
\end{eqnarray}
where $\dot{\phantom{x}}={\rm d}/{\rm d}\tau$, with $\tau $ an affine parameter distinct from the proper time.
The upper photon would arrive at Earth with final conditions $(\tau=0)$ $t =0$, $\chi =\chi _{\text{L}}$, $\varphi =\pi$, $\dot{t}=1$, $\dot{\chi} =\cos\beta/a_{0}$, and $\dot{\varphi}=\sin\beta/a_{0}$, where we have used the fact that the physical angle $\alpha$ coincides with the coordinate angle, i.e., $\tan\alpha = \left\vert r_{\text{E}}\text{d}\varphi (r_{\text{E}})/\text{d%
}r\right\vert =\left\vert \tanh \chi _{\text{L}}\text{d}\varphi
(\chi _{\text{L}})/\text{d}\chi \right\vert$. Then, a straightforward integration of  Eqs.~(\ref{geqFLRW1}) and~(\ref{geqFLRW3}), together with the normalization condition $\dot{s}_\text{FLRW} =\varepsilon$ ($\varepsilon=0$ for null geodesics), yields
\begin{eqnarray}
   && \dot{t}=\frac{a_{0}}{a(t)},\quad \dot{\varphi}=\frac{a_{0}\sinh\chi _{\text{PE}}}{
a^{2}\sinh ^{2}\chi },\\
&&\varphi =\pi -\arcsin \frac{\tanh \chi_{\text{PE}}}{\tanh\chi}+\beta , \label{sol1}
\end{eqnarray}
from which a relation between the variables $\chi$ and $a$ can be derived,
\begin{equation}
\frac{\text{d}a}{af(a)}=\frac{\text{d}\chi }{\sqrt{1-\chi_{\text{PE}}^{2}/\sinh ^{2}\chi }}, \label{chi-a}
\end{equation}
where $\chi _{\text{PE}}$ and $\beta$ are constants defined by $\sinh\chi _{\text{PE}}=\sinh\chi _{\text{L}}\sin\beta$ and $\sin \beta =\cosh\chi_{\text{L}}\tan\alpha$. The integration of the right-hand side of Eq.~(\ref{chi-a}) can be analytically carried out leading to
\begin{equation}
\int_{a_{\text{Sch\"{u}E}}}^{a_{0}}\frac{\text{d}a}{af(a)}=\text{arccosh}
\frac{\cosh \chi _{\text{L}}}{\cosh \chi _{\text{PE}}}-\text{arccosh}\frac{
\cosh \chi _{\text{Sch\"{u}}}}{\cosh \chi _{\text{PE}}},  \label{aschuE}
\end{equation}
from which the value of $a_{\text{Sch\"{u}E}}=a(t_{\text{Sch\"{u}E}})$ is calculated by numerical integration. The corresponding value of $t_{\text{Sch\"{u}E}}$ at which the upper photon emerges from the vacuole is then calculated by numerical integration of the Friedmann equation (\ref{FLRWeq}), i.e.,
\begin{equation}
t_{\text{Sch\"{u}E}}=\int_{a_{0}}^{a_{\text{Sch\"{u}E}}}\frac{\text{d}a}{%
f(a)}.  \label{tschuE}
\end{equation}
Similar formulas apply in the case of the lower photon, with $\pi$ replaced by $-\pi$ and $\alpha$ replaced by $-\alpha^\prime$ regarding the Earth position.

However, one can proceed differently and calculate $t_{\text{Sch\"{u}E}}^\prime$, at which the lower photon emerges from the vacuole, in terms of $t_{\text{Sch\"{u}E}}$ through an approximate formula. Subtracting the right and the left-hand sides of Eq.~(\ref{aschuE}) from the ones which correspond to the lower photon, then approximating in the limit of small $\alpha$ and $\alpha^\prime$, one gets
\begin{eqnarray}
\int_{t_{\text{Sch\"{u}E}}}^{t_{\text{Sch\"{u}E}}^{\prime }}\frac{\text{d}t}{a(t)}&\simeq &\frac{1}{2}\left( \coth\chi_{\text{Sch\"{u}}}-\coth\chi_{\text{L}}\right) \sinh ^{2}\chi_{\text{L}}\nonumber\\
    &&\times\cosh ^{2}\chi_{\text{L}}(\alpha ^{2}-\alpha ^{\prime 2}).\label{TD_exit}
\end{eqnarray}
Now, to approximate the left-hand side of Eq.~(\ref{TD_exit}), one can use the fact that $a(t)$ vary appreciably only on cosmological time intervals. The final result is
\begin{eqnarray}
\Delta t_{\text{Sch\"{u}E}}&\simeq &\frac{1}{8}a_{\text{Sch\"{u}E}}\left( \coth\chi _{\text{Sch\"{u}}}-\coth\chi _{
\text{L}}\right)\sinh ^{2}(2\chi _{\text{L}}) \nonumber\\
&&\times(\alpha^{2}-\alpha^{\prime2}).\label{TD_exit_final}  
\end{eqnarray}
where we have put $\Delta t_{\text{Sch\"{u}E}}=t_{\text{Sch\"{u}E}}^{\prime }-t_{\text{Sch\"{u}E}}$.

\subsection{Integrating null geodesic equations inside the vacuole}
In this region described by the SdS metric, the main task is to determine the scale factors $a _{\text{Sch\"{u}S}}$ and $a _{\text{Sch\"{u}S}}^\prime$ and their corresponding times $t _{\text{Sch\"{u}S}}$ and $t _{\text{Sch\"{u}S}}^\prime$, at which the upper and lower photons enter the vacuole. This is achieved by integrating geodesic equations backward from the known final conditions at the exit points from the vacuole.

The time required for the upper photon to traverse the vacuole is obtained by integration of Eq.~(\ref{timematch}),
\begin{equation}
T_{\text{Sch\"{u}E}}-T_{\text{Sch\"{u}S}}=\cosh \chi _{\text{Sch\"{u}}
}\int_{a_{\text{Sch\"{u}S}}}^{a_{\text{Sch\"{u}E}}}\frac{
\text{d}a}{B_{\text{Sch\"{u}}}(a)f(a)},\label{travel1}
\end{equation}
using the Friedmann equation (\ref{FLRWeq}). Another expression for this travel time can be obtained by making use of the well-known equation
\begin{equation}
\text{$d$}T=\pm \frac{\text{d}r}{v(r)},\quad v(r)=B(r)\sqrt{1-J^{2}\frac{B(r)}{r^{2}}}, \label{TDSdS1}
\end{equation}
which follows directly from the first and second partially integrated SdS null geodesic equations, i.e.,
\begin{equation}
\dot{T}\,=\frac{1}{B(r)},\quad \dot{r}=\pm \sqrt{1-J^{2}\frac{B(r)}{r^{2}}},\quad \dot{\varphi}=\frac{J}{r^{2}},  \label{sol2}
\end{equation}
where $J$ is a constant of motion defined by $J=r_{\text{P}}/\sqrt{B(r_{\text{P}})}$ with $r_\text{P}$ the perilens given approximately by \cite{Schu1} $r_{\text{P}}\simeq r_{\text{Sch\"{u}E}}\sin \gamma _{\text{SdS}}-GM$ with $\tan \gamma _{\text{SdS}}=\left\vert r_{\text{Sch\"{u}E}}\text{d}\varphi (r_{\text{Sch\"{u}E}})/\text{d}r\right\vert$ calculated using the inverse Jacobian of~(\ref{J}) and $r_{\text{Sch\"{u}E}}=a_{\text{Sch\"{u}E}}\sinh \chi_{\text{Sch\"{u}}}$. The result is obtained by integrating Eq.~(\ref{sol2}), i.e.,
\begin{equation}
T_{\text{Sch\"{u}E}}-T_{\text{Sch\"{u}S}}=\left( \int_{r_{\text{P}}}^{r_{
\text{Sch\"{u}E}}}+\int_{r_{\text{P}}}^{r_{\text{Sch\"{u}S}}}\right) \frac{
\text{d}r}{v(r)},  \label{travel2}
\end{equation}
using the fact that $r$ decreases over $[r_{\text{Sch\"{u}S}},r_{\text{P}}]$ while it increases over $[r_{\text{P}},r_{\text{Sch\"{u}E}}]$. Equating the two right sides of Eqs.~(\ref{travel1}) and~(\ref{travel2}), one obtains
\begin{eqnarray}
 \left( \int_{r_{\text{P}}}^{r_{\text{Sch\"{u}E}}}+\int_{r_{\text{P}}}^{r_{\text{Sch\"{u}S}}}\right) 
\frac{\text{d}r}{v(r)}&=&\int_{a_{\text{Sch\"{u}S}}}^{a_{\text{Sch\"{u}E}}}\frac{\text{d}a}{B_{\text{Sch\"{u}}}(a)f(a)}\nonumber\\
&&\times\cosh \chi_{\text{Sch\"{u}}},\label{TD_entry}
\end{eqnarray}
which enables us to obtain $a _{\text{Sch\"{u}S}}$ by numerical integration, with $r_{\text{Sch\"{u}S}}=a_{\text{Sch\"{u}S}}\sinh \chi_{\text{Sch\"{u}}}$. Then one has to integrate numerically the Friedmann equation (\ref{FLRWeq}) to obtain $t_{\text{Sch\"{u}S}}$, i.e.,
\begin{equation}
t_{\text{Sch\"{u}S}}=\int_{a_{0}}^{a_{\text{Sch\"{u}S}}}\frac{\text{d}a}{f(a)}.  \label{tschuS}
\end{equation}

Similar formulas are applied for the lower photon leading to the calculation of the scale factor $a _{\text{Sch\"{u}S}}^\prime$ and $t _{\text{Sch\"{u}S}}^\prime$. This also can be determined by difference with $t _{\text{Sch\"{u}S}}$ via an approximate analytical expression. Subtracting the right and the left-hand side of Eq.~(\ref{travel1}) from the ones that correspond to the lower photon, and taking into account that $B _{\text{Sch\"{u}}}$ is only significant on cosmological timescales, one gets
\begin{equation}
\Delta T_{\text{Sch\"{u}E}}-\Delta T_{\text{Sch\"{u}S}} \simeq \left( \frac{\Delta t_{
\text{Sch\"{u}E}}}{B_{\text{Sch\"{u}E}}}-\frac{\Delta t_{\text{Sch\"{u}S}}}{B_{\text{Sch\"{u}S}}}\right)\cosh \chi _{\text{Sch\"{u}}}  ,  \label{dif_travel1}
\end{equation}
with $\Delta T_{\text{Sch\"{u}E}}=T_{\text{Sch\"{u}E}}^{\prime }-T_{\text{Sch\"{u}E}}$, $\Delta T_{\text{Sch\"{u}S}}=T_{\text{Sch\"{u}S}}^{\prime }-T_{\text{Sch\"{u}S}}$, $\Delta t_{\text{Sch\"{u}S}}=t_{\text{Sch\"{u}S}}^{\prime }-t_{\text{Sch\"{u}S}}$, $B_{\text{Sch\"{u}E}}=B_{\text{Sch\"{u}}}(t_{\text{Sch\"{u}E}})$ and $B_{\text{Sch\"{u}S}}=B_{\text{Sch\"{u}}}(t_{\text{Sch\"{u}S}})$. This expresses the difference in the travel times between both photons inside the vacuole, which can be otherwise obtained by the use of Eq.~(\ref{TDSdS1}), i.e.,
\begin{eqnarray}
\Delta T_{\text{Sch\"{u}E}}-\Delta T_{\text{Sch\"{u}S}}&=&\left(\int_{r_{\text{Sch\"{u}E}}}^{r_{\text{Sch\"{u}E}}^{\prime }}+\int_{r_{\text{Sch\"{u}S}}}^{r_{\text{Sch\"{u}S}}^{\prime}}\right)\frac{\text{d}r}{v(r)}\nonumber\\&&-\Delta T_{\text{SdS}},\label{dif_travel2}
\end{eqnarray}
where we have divided the integrals to produce the following expression
\begin{eqnarray}
\Delta T_{\text{SdS}}&=&\left( \int_{r_{\text{P}}}^{r_{\text{Sch\"{u}E}
}}+\int_{r_{\text{P}}}^{r_{\text{Sch\"{u}S}}}\right) \frac{\text{d}r}{v(r)}\nonumber\\
&&-\left( \int_{r_{\text{P}}^{\prime }}^{r_{\text{Sch\"{u}E}}}+\int_{r_{\text{P}}^{\prime}}^{r_{\text{Sch\"{u}S}}}\right) \frac{\text{d}r}{v^{\prime }(r)}.  \label{TD_SdS1}
\end{eqnarray}
This can be directly compared to an expression previously used in the calculation of the time delay within the framework of the SdS solution \cite{Schu2}, i.e.,
\begin{eqnarray}
\Delta T_{\text{SdS}} &\simeq &\frac{1}{2}\left( r_{\text{P}}^{\prime 2}-r_{
\text{P}}^{2}\right) \left( \frac{1}{r_{\text{Sch\"{u}E}}}+\frac{1}{r_{
\text{Sch\"{u}S}}}\right)\nonumber\nonumber\\&&+4GM\ln \frac{r_{\text{P}}^{\prime}}{r_{\text{P}}}-\frac{3}{2}G^2M^2\left( \frac{1}{r_{\text{P}}^{2}}-\frac{1}{r_{\text{P}}^{\prime 2}}
\right) \sqrt{\frac{3}{\Lambda }}  \nonumber \\
&&\times \left( 
\text{arctanh} \sqrt{\frac{\Lambda }{3}r_{\text{Sch\"{u}E}}^2} +
\text{arctanh} \sqrt{\frac{\Lambda }{3}r_{\text{Sch\"{u}S}}^2} 
\right) .\nonumber\\&&\phantom{}  \label{TD_SdS2}
\end{eqnarray}
Additionally, given that the lengths and timescales under consideration are shorter than cosmological scales, the last term on the right-hand side of Eq.~(\ref{dif_travel2}) can be approximated as
\begin{eqnarray}
\left( \int_{r_{\text{Sch\"{u}E}}}^{r_{\text{Sch\"{u}E}}^{\prime }}+\int_{r_{
\text{Sch\"{u}S}}}^{r_{\text{Sch\"{u}S}}^{\prime }}\right) \frac{\text{d}r}{v(r)} &\simeq &\sinh \chi _{\text{Sch\"{u}}}\left( f_{\text{Sch\"{u}E}}\frac{\Delta t_{\text{Sch\"{u}E}}}{v{}_{\text{Sch\"{u}E}}}\right . \nonumber\\ 
&& \left .+f_{\text{Sch\"{u}S}}
\frac{\Delta t_{\text{Sch\"{u}S}}}{v{}_{\text{Sch\"{u}S}}}\right),\label{last_term}    
\end{eqnarray}
using the matching condition (\ref{match}) as well as the Friedmann equation (\ref{FLRWeq}), with $v{}_{\text{Sch\"{u}E}}=f(r_{\text{Sch\"{u}E}})$, $v{}_{\text{Sch\"{u}S}%
}=v(r_{\text{Sch\"{u}S}})$, $f_{\text{Sch\"{u}E}}=f(a_{\text{Sch\"{u}E}})$ and $f_{\text{Sch\"{u}S}}=f(a_{\text{Sch\"{u}S}})$. Replacing (\ref{last_term}) together with (\ref{TD_SdS2}) in Eq.~(\ref{dif_travel2}) and equating its right-hand side with the one of Eq.~(\ref{dif_travel1}), we finally arrive at
 \begin{widetext}
 \begin{equation}
 \Delta t_{\text{Sch\"{u}S}}\simeq\frac{\Delta T_{\text{SdS}}+\left( \dfrac{\cosh \chi _{\text{Sch\"{u}}}}{B_{\text{Sch\"{u}E}}}-\dfrac{f_{\text{Sch\"{u}E}}\sinh \chi _{\text{Sch\"{u}}}}{v{}_{\text{Sch\"{u}E}}}\right)  \Delta t_{\text{Sch\"{u}E}} }{\dfrac{\cosh \chi _{\text{Sch\"{u}}}}{B_{\text{Sch\"{u}S}}}+\dfrac{f_{\text{Sch\"{u}S}}\sinh \chi _{\text{Sch\"{u}}}}{v{}_{\text{Sch\"{u}S}}}}. \label{TD_entry_final}
\end{equation}
\end{widetext}

The determination of $t_{\text{Sch\"{u}S}}$ allows us to compute the polar angle $\varphi_{\text{Sch\"{u}S}}$ at which the photon arrives on the boundary of the vacuole, a quantity required for the subsequent analysis in Sec.~\ref{subsecC}. This is achieved by employing the well-known equation
\begin{equation}
\text{$d$}\varphi=\pm \frac{\text{d}r}{u(r)},\quad u(r)=r\sqrt{\frac{r^{2}}{J^{2}}-B(r)}, \label{defSdS1}
\end{equation}
which follows immediately from the second and third partially integrated SdS null geodesic equations in (\ref{sol2}). Integrating Eq.~(\ref{defSdS1}) yields
\begin{equation}
\varphi_{\text{Sch\"{u}E}}-\varphi_{\text{Sch\"{u}S}}=\left( \int_{r_{\text{P}}}^{r_{
\text{Sch\"{u}E}}}+\int_{r_{\text{P}}}^{r_{\text{Sch\"{u}S}}}\right) \frac{
\text{d}r}{v(r)},  \label{varphiSchS1}
\end{equation}
where we account for the fact that $\varphi$ increases both as the photon approaches the lens and as it recedes from it. The right-hand side of Eq.~(\ref{varphiSchS1}) can be evaluated analytically to leading order in the ratio of Schwarzschild radius to perilens, $2GM/r_\text{P}$, resulting in
\begin{eqnarray}
\varphi _{\text{Sch\"{u}S}} &\simeq &\varphi _{\text{Sch\"{u}E}}-\pi
+\arcsin \frac{r_{\text{P}}}{r_{\text{Sch\"{u}E}}}+\arcsin \frac{r_{\text{P}}}{r_{\text{Sch\"{u}S}}}  \nonumber \\
&&-\frac{GM}{r_{\text{P}}}\left( \sqrt{1-\frac{r_{\text{P}}^{2}}{r_{
\text{Sch\"{u}E}}^{2}}}+
\sqrt{1-\frac{r_{\text{P}}^{2}}{r_{\text{Sch\"{u}S}}^{2}}}\right. \nonumber \\
&&\left. +\sqrt{\frac{r_{\text{Sch\"{u}E}}-r_{\text{P}}}{r_{\text{Sch\"{u}E}}+r_{\text{P}}}}+\sqrt{\dfrac{r_{\text{Sch\"{u}S}}-r_{\text{P}}}{
r_{\text{Sch\"{u}S}}+r_{\text{P}}}}\right),\label{varphiSchS2}
\end{eqnarray}
with $\varphi _{\text{Sch\"{u}E}}=\varphi(t _{\text{Sch\"{u}E}})=\varphi(\chi _{\text{Sch\"{u}}})$ given by Eq.~(\ref{sol1}).

\subsection{\label{subsecC}Integrating null geodesic equations between vacuole and source}
The motion of photons in this region is governed by the same FLRW geodesic equations, (\ref{geqFLRW1}) and (\ref{geqFLRW3}), treated in Sec.~\ref{subsecA}. One must integrate them in order to obtain the scale factor $a_\text{S}$ and its corresponding emission time $t_\text{S}$ of the upper photon at the source S, taking into account the final conditions at the entry into the vacuole, i.e., $t=t_{\text{Sch\"{u}S}}$, $\chi=\chi _{\text{Sch\"{u}}}$, $\varphi =\varphi_{\text{Sch\"{u}S}}$, $\dot{t}=\dot{t}_{\text{Sch\"{u}S}}$, $\dot{\chi}=\dot{\chi}_{\text{Sch\"{u}S}}$, and $\dot{\varphi}=\dot{\varphi}_{\text{Sch\"{u}S}}$, which are calculated using the Jacobian (\ref{J}). We obtain
\begin{eqnarray}
 &&\dot{t}=\frac{E}{a(t)},\quad \dot{\varphi}=\frac{J}{a^{2}\sinh^{2}\chi},\label{sol3_0}\\
 && \varphi =\varphi _{\text{Sch\"{u}S}}+\arcsin \frac{\tanh\chi_{\text{PS}}}{\tanh\chi}-\gamma,  \label{sol3}   
\end{eqnarray}
where $E$, $\chi _{\text{PS}}$, and $\gamma$ are constants related to each other by $E=a_{\text{Sch\"{u}S}}\dot{t}_{\text{Sch\"{u}S}}$, $\sinh\chi_{\text{PS}}=J/E$,  and $\sin\gamma =\tanh\chi_{\text{PS}}/\tanh\chi_{\text{Sch\"{u}}}$. Hence, the source angular position $\varphi _\text{S}$ corresponding to the comoving distance $\chi _{%
\text{LS}}$ (\ref{chiLS}) is calculated using Eq.~(\ref{sol3}), i.e.,
\begin{equation}
-\varphi _{\text{S}}=-\varphi _{\text{Sch\"{u}S}}-\arcsin \frac{\tanh\chi_{\text{PS}}}{\tanh\chi_\text{LS}}+\gamma,  \label{phiS}
\end{equation}
where $\varphi _{\text{Sch\"{u}S}}$ is given by (\ref{varphiSchS2}).

In a similar manner to the upper photon, analogous formulas are obtained for the lower photon, with the only difference being that the constant $J^\prime$ differs by a minus sign, $J^{\prime}=-r_{\text{P}}^{\prime}/\sqrt{B(r_{\text{P}}^{\prime})}$ with $r_{\text{P}}^{\prime}\simeq r_{\text{Sch\"{u}E}}^{\prime}\sin \gamma _{\text{SdS}}^{\prime}-GM$ and $\tan \gamma _{\text{SdS}}^{\prime}=\left\vert r_{\text{Sch\"{u}E}}^{\prime}\text{d}\varphi (r_{\text{Sch\"{u}E}}^{\prime})/\text{d}r\right\vert$. The expression of the source angular position $\varphi _{\text{S}}^\prime$, which corresponds to the same comoving distance $\chi _{%
\text{LS}}$, is then
\begin{equation}
-\varphi _{\text{S}}^\prime=-\varphi _{\text{Sch\"{u}S}}^\prime+\arcsin \frac{\tanh\chi
_{\text{PS}}^\prime}{\tanh\chi_{\text{LS}}}-\gamma^\prime,
\label{phiS_prime}
\end{equation}
where $\chi_{\text{PS}}^\prime$ is a constant related to $J^\prime$ and the constant of motion $E^\prime=a_{\text{Sch\"{u}S}}^\prime\dot{t}_{\text{Sch\"{u}S}}^\prime$ by $\sinh\chi_{\text{PS}}^\prime=J^\prime/E^\prime$, and $\sin\gamma ^\prime=\tanh\chi_{\text{PS}}^\prime/\tanh\chi_{\text{Sch\"{u}}}$.

These angles $\varphi _{\text{S}}$ and $\varphi _{\text{S}}^{\prime }$ must coincide, as both photons originate from the same source. Among all the parameters involved, adjusting the lens mass appears to be the only effective means to equate them. Once the accurate mass is determined, the calculation of the time delay can proceed.

As before, an equation similar to (\ref{chi-a}) linking $\chi$ to $a$ can be derived from  Eqs.~(\ref{sol3_0}) and (\ref{sol3}),
\begin{equation}
\frac{\text{d}a}{af(a)}=-\frac{\text{d}\chi }{\sqrt{1-\sinh^{2}\chi_{\text{PS}}/\sinh ^{2}\chi }}, \label{chi-aS}
\end{equation}
where the minus sign reflects a decreasing function over time in this region. Then integrating between $\chi _{\text{LS}}$ and $\chi _{\text{Sch\"{u}}}$ leads to
\begin{equation}
\int_{a_{\text{S}}}^{a_{\text{Sch\"{u}S}}}\frac{\text{d}a}{af(a)}= \text{arccosh}\frac{\cosh\chi_{\text{LS}}}{\cosh\chi_{\text{PS}}}- \text{arccosh} \frac{\cosh\chi_{\text{Sch\"{u}}}}{\cosh\chi_{\text{PS}}},  \label{aS}
\end{equation}
which can be numerically solved to calculate $a_{\text{S}}$. Then, $t_\text{S}$ is calculated through the Friedmann equation as
\begin{equation}
t_{\text{S}}=\int_{a_{0}}^{a_{\text{S}}}\frac{\text{d}a}{f(a)}.  \label{tSlower}
\end{equation}
Similar formulas are applied regarding the lower photon, replacing $a_{\text{Sch\"{u}S}}$ by $a_{\text{Sch\"{u}S}}^{\prime }$ and $\chi _{\text{PS}}$ by $\chi _{\text{PS}}^{\prime }$.

Now, let us proceed differently to calculate the time delay via an approximate expression. Subtracting the two sides of (\ref{aS}) from the ones of the lower photon, and approximating in the limit of small $\varphi _{\text{Sch\"{u}S}}-\varphi _{\text{S}}$ and $|\varphi _{\text{Sch\"{u}S}}^{\prime }-\varphi _{\text{S}}|$ ($\sim 10^{-4}$), one gets
\begin{widetext}
\begin{equation}
\left(\int_{t_{\text{Sch\"{u}S}}}^{t_{\text{Sch\"{u}S}}^{\prime }}-\int_{t_{\text{S}}}^{t_{\text{S}}^{\prime }}\right)\frac{\text{d}t}{a(t)}
\simeq \frac{1}{2}\frac{(\varphi _{\text{Sch\"{u}S}}^{\prime }-\varphi _{
\text{S}})^{2}-(\varphi _{\text{Sch\"{u}S}}-\varphi _{\text{S}%
})^{2}}{\coth\chi_{\text{Sch\"{u}}}-\coth\chi_{\text{LS}}}.\label{TDS}
\end{equation}
Approximating the left-hand side of (\ref{TDS}) using the fact that the scale factor $a(t)$ varies noticeably only over cosmological timescales, one obtains finally the time delay formula
\begin{equation}
\Delta t\simeq a_{\text{S}}\left[ \frac{\Delta t_{\text{Sch\"{u}S}}}{a_{\text{Sch\"{u}S}}}-\frac{1}{2}\frac{(\varphi _{\text{Sch\"{u}S}}^{\prime }-\varphi _{\text{S}})^{2}-(\varphi _{\text{Sch\"{u}S}}-\varphi _{\text{S}})^{2}}{\coth\chi_{\text{Sch\"{u}}}-\coth\chi_{\text{LS}}}\right].  \label{TD}
\end{equation}
\end{widetext}
\section{\label{sec4}The lensed quasar SDSS J1004+4112}
The results are now applied to the lensed quasar SDSS J1004+4112 discovered by Inada et al. \cite{Inada}. This system clearly exhibits a strong gravitational lensing caused by a galaxy cluster at a redshift $z_\text{L}=0.68$, producing five images of a background quasar at redshift $z_\text{S}=1.734$. The lens is assumed to be symmetric, despite the fact that the presence of five images shows it is not. Here, we only consider the very large angular separation of about 15 arcseconds between images C and D, with $\alpha=5^{\prime\prime}\pm10\%$ (D) and $\alpha^\prime=10^{\prime\prime}\pm10\%$ (C).
\begin{table}[htbp]
\caption{Galaxy cluster mass $M$, position angle $-\varphi_\text{S}$, and time delay $\Delta t$ for the observed image pair (D,C) of the lensed quasar SDSS J1004+4112 in hyperbolic Einstein-Straus--de Sitter spacetime ($k=-1$). The cosmic scale factor is fixed in the value $a_0=5\,\text{am}$ ($\Omega_{k0}=-0.04$).}\label{tab1}
\begin{ruledtabular}
\begin{tabular}{lccr}
$(\Lambda ,\alpha ,\alpha ^{\prime })$ & $M[10^{13}M_{\odot }]$ & $-\varphi_\text{S}[^{\prime \prime }]$& $\Delta t[\text{yr}]$\\ \hline
$(+,+,+)$ & $2.208632$& $9.89304$& $12.1077$\\ 
$(+,+,\pm 0)$ & $2.007476$& $8.10108$& $9.40788$\\ 
$(+,+,-)$ & $1.806348$& $6.30938$& $6.92009$\\ 
$(+,\pm0,+)$& $2.008189$& $10.7857$& $12.6577$\\ 
$(+,\pm 0,\pm 0)$& $1.8253007$& $8.99358$& $10.0078$\\ 
$(+,\pm 0,-)$ & $1.6424348$& $7.20165$& $7.56259$\\ 
$(+,-,+)$ & $1.8077$& $11.6787$& $13.0953$\\ 
$(+,-,\pm 0)$ & $1.6430808$& $9.88629$& $10.5029$\\ 
$(+,-,-)$ & $1.478481$& $8.09413$& $8.10762$\\ 
$(\pm 0,+,+)$ & $2.24554$& $11.0019$& $12.0502$\\ 
$(\pm 0,+,\pm 0)$& $2.041158$& $9.00591$& $9.3712$\\ 
$(\pm 0,+,-)$ & $1.836798$& $7.01012$& $6.89681$\\ 
$(\pm 0,\pm 0,+)$ & $2.041581$& $11.9976$& $12.5832$\\ 
$(\pm 0,\pm 0,\pm 0)$ & $1.855775$& $10.0015$& $9.96047$\\ 
$(\pm 0,\pm 0,-)$ & $1.6699799$& $8.00555$& $7.53374$\\ 
$(\pm 0,-,+)$ & $1.8375975$& $12.9934$& $12.9971$\\ 
$(\pm 0,-,\pm 0)$ & $1.670366$& $10.9972$& $10.4397$\\ 
$(\pm 0,-,-)$ & $1.503141$& $9.0011$& $8.06935$\\ 
$(-,+,+)$ & $2.220931$& $11.6782$& $11.7829$\\ 
$(-,+,\pm 0)$ & $2.01884$& $9.55814$& $9.16816$\\ 
$(-,+,-)$ & $1.8167686$& $7.43828$& $6.74997$\\ 
$(-,\pm 0,+)$ & $2.019142$& $12.7363$& $12.2957$\\ 
$(-,\pm 0,\pm 0)$ & $1.835426$& $10.6162$& $9.7396$\\ 
$(-,\pm 0,-)$ & $1.65171415$& $8.49622$& $7.37091$\\ 
$(-,-,+)$ & $1.817342$& $13.7945$& $12.6882$\\ 
$(-,-,\pm 0)$ & $1.625673$& $12.1218$& $9.96458$\\ 
$(-,-,-)$ & $1.4866455$& $9.55423$& $7.89049$\\ 
\end{tabular}
\end{ruledtabular}
\end{table}
\begin{table}[htbp]
\caption{Variation of the galaxy cluster mass $M$, the position angle $-\varphi_\text{S}$, and the time delay $\Delta t$ versus the current cosmic scale factor $a_0$ for the observed image pair (D,C) of the lensed quasar SDSS J1004+4112 in hyperbolic Einstein-Straus--de Sitter spacetime ($k=-1$). The cosmological constant $\Lambda$ as well as the angles $\alpha$ and $\alpha^\prime$ are fixed in their central values.}\label{tab2}
\begin{ruledtabular}
\begin{tabular}{lccr}
$a_{0}[\text{am}]$& $M[10^{13}M_{\odot }]$ & $-\varphi _\text{S}[^{\prime \prime}]$& $\Delta t[\text{yr}]$\\ \hline
$2.1$ & $2.016437$& $10.0961$& $11.301$\\
$2.2$ & $1.99955$ &$10.0913$ &$11.1473$ \\ 
$2.3$ & $1.984515$ & $10.0854$ & $11.01$\\ 
$2.4$ & $1.971193$ & $10.0792$ & $10.8918$ \\
$2.5$ & $1.9593693$ & $10.0731$ & $10.7893$ \\ 
$2.6$ & $1.948855$ & $10.0673$ & $10.6999$ \\
$2.7$ & $1.939484$& $10.0617$ & $10.6215$\\ 
$2.8$ & $1.931099$& $10.0565$& $10.5522$\\ 
$2.9$ & $1.923578$& $10.0517$& $10.4907$\\ 
$3.0$ & $1.916802$& $10.0472$& $10.4359$\\ 
$3.2$ & $1.905132$ & $10.0392$ & $10.3425$ \\
$3.4$ & $1.895497$ & $10.0322$ & $10.2663$ \\
$3.6$ & $1.88745$ & $10.0263$ & $10.2034$ \\
$3.8$ & $1.880665$ & $10.0212$ & $10.1507$ \\
$4.0$ & $1.874885$ & $10.0167$ & $10.1061$ \\ 
$4.5$ & $1.863721$ & $10.0079$ & $10.0207$ \\
$5.0$ & $1.855775$ & $10.0015$ & $9.96047$ \\ 
$6.0$ & $1.845481$ & $9.99297$ & $9.88309$ \\ 
$7.0$ & $1.839308$ & $9.98776$ & $9.83703$ \\ 
$8.0$ & $1.835316$ & $9.98435$ & $9.80736$ \\ 
$9.0$ & $1.832585$ & $9.98200$ & $9.78712$ \\ 
$10.0$ & $1.8306345$ & $9.98031$ & $9.7727$\\ 
$11.0$ & $1.829193$ & $9.97906$ & $9.76206$ \\
$12.0$ & $1.828098$ & $9.97811$ & $9.75398$ \\ 
$13.0$ & $1.8272465$ & $9.97737$ & $9.7477$ \\
$14.0$ & $1.8265714$ & $9.97678$ & $9.74273$ \\
$15.0$ & $1.8260268$ & $9.9763$ & $9.73872$ \\ 
$16.0$ & $1.825581$ & $9.97591$ & $9.73544$ \\
$17.0$ & $1.8252119$ & $9.97559$ & $9.73272$ \\
$18.0$ & $1.8249025$ & $9.97532$ & $9.73044$ \\
$19.0$ & $1.8246409$ & $9.97509$ & $9.72852$ \\
$20.0$ & $1.824417$ & $9.97489$ & $9.72688$ \\
$21.0$ & $1.824225$ & $9.97472$ & $9.72546$ \\
$22.0$ & $1.8240586$ & $9.97457$ & $9.72424$ \\
$23.0$ & $1.8239134$ & $9.97445$ & $9.72317$ \\
$24.0$ & $1.82378576$ & $9.97433$ & $9.72224$ \\
$25.0$ & $1.8236734$ & $9.97424$ & $9.72141$ \\
$26.0$ & $1.823574$ & $9.97415$ & $9.72068$ \\
$27.0$ & $1.823485$ & $9.97407$ & $9.72002$ \\
$28.0$ & $1.823405$ & $9.974$ & $9.71944$ \\
$29.0$ & $1.8233336$ & $9.97394$ & $9.71891$ \\
$30.0$ & $1.823269$ & $9.97388$ & $9.71844$ \\
$31.0$ & $1.823211$ & $9.97383$ & $9.71801$ \\
$32.0$ &  $1.823158$ & $9.97378$ & $9.71762$ \\
$33.0$ & $1.82311$ & $9.97374$ & $9.71727$ \\
$34.0$ & $1.823066$ & $9.9737$ & $9.71695$ \\
$35.0$ & $1.8230256$ & $9.97367$ & $9.71665$ \\
$36.0$ & $1.8229884$ & $9.97363$ & $9.71638$ \\
$37.0$ & $1.8229545$ & $9.9736$ & $9.71613$ \\
$38.0$ & $1.8229234$ & $9.97358$ & $9.7159$ \\
$39.0$ & $1.822894$ & $9.97355$ & $9.71569$ \\
$40.0$ & $1.8228676$ & $9.97353$ & $9.71549$
\end{tabular}
\end{ruledtabular}
\end{table}

For numerical convenience, we use a system of astro-units introduced by Schücker in Ref.~\cite{Schu1}, in which time, distances, and masses are measured in astroseconds (as), astrometers (am) and astrograms (ag), respectively, as follows:
\begin{eqnarray}
&&1\,\text{as}=4.34\times 10^{17}\,\text{s}=13.8\,\text{Gyr}, \\
&&1\,\text{am}=1.30\times 10^{26}\,\text{m}=4221\,\text{Mpc}, \\
&&1\,\text{ag}=6.99\times 10^{51}\,\text{kg}=3.52\times 10^{21}M_{\odot }.
\label{unit}
\end{eqnarray}
In this astro-unit system, the fundamental constants take the values: $c=1\,\text{am}\,\text{as}^{-1}$, $8\pi G=1\,\text{am}^{3}\,\text{as}^{-2}\,\text{ag}^{-1}$, and $H_{0}=1\,\text{as}^{-1}$ ($=71\,\text{km}\,\text{s}^{-1}\,\text{Mpc}^{-1}$), with $H_{0}$ the Hubble constant involved in the Hubble law $\text{d}a/\text{d}t=H_{0}a(t)$. Substituting this into the Friedmann equation  (\ref{FLRWeq}) yields $A=\rho _{0}a_{0}^{3}/3$ with the current dust density $\rho _{0}=3-\Lambda +3\Omega _{k0}$ and current curvature density $\Omega _{k0}=k/a_{0}^{2}=-1/a_{0}^{2}$.  We use throughout this work the same cosmological constant adopted in \cite{Schu1,Guen1,Guen2}, $\Lambda=1.36\times10^{-52}\,\text{m}^{-2}=0.77\times 3\,\text{am}^{-2}$, with an accuracy of $20\%$. To guarantee a definite positive dust density $\rho _{0}>0$, the current scale factor $a_0$ must satisfy $a_{0}>1/\sqrt{1-\Lambda /3}\simeq 2.08514\,\text{am}$. This lower bound translates into a constraint on the current curvature density, $|\Omega_{k0}|<0.23$, ensuring that the cosmological parameters are physically consistent with the Friedmann equation in the adopted unit system. The comoving distances $\chi_\text{L}$ and $\chi_\text{S}$, and $\chi_\text{LS}$ are calculated from (\ref{comovdist}) and (\ref{chiLS}).

\begin{figure}[htbp]
\centering
\begin{tikzpicture}
\begin{axis} [axis lines=middle, width=8.7cm, height=5cm, xlabel={$a_0$[am]}, ylabel={$M[10^{13} M_{\odot }]$}, xmin=2.1, xmax=40, ymin=1.79, ymax=2.02, grid=both, xlabel style={at={(axis description cs:0.5,-0.1)}, anchor=north}, ylabel style={at={(axis description cs:-0.09,0.5)}, anchor=south,rotate=90}]

 \addplot[smooth, thick, blue] coordinates {(2.1,2.016437) (2.2,1.99955) (2.3,1.984515) (2.4,1.971193) (2.5,1.9593693) (2.6,1.948855) (2.7,1.9394840) (2.8,1.931099) (2.9,1.923578) (3.0,1.916802) (3.2,1.905132) (3.4,1.895497) (3.6,1.88745) (3.8,1.880665) (4.0,1.874885) (4.5,1.863721) (5.0,1.855775) (6.0,1.845481) (7.0,1.839308) (8.0,1.835316) (9.0,1.832585) (10.0,1.8306345) (11.0,1.829193) (12.0,1.828098) (13.0,1.8272465) (14.0,1.8265714) (15.0,1.8260268) (16.0,1.825581) (17.0,1.8252119) (18.0,1.8249025) (19.0,1.8246409) (20.0,1.824417) (21,1.824225) (22,1.8240586) (23,1.8239134) (24,1.82378576) (25,1.8236734) (26,1.823574) (27,1.823485) (28,1.823405) (29,1.8233336) (30.0,1.823269) (31,1.823211) (32,1.823158) (33,1.82311) (34,1.823066) (35,1.8230256) (36,1.8229884) (37,1.8229545) (38,1.8229234) (39,1.822894) (40,1.8228676)};
 
\addplot[only marks, mark=*, red, mark size=0.7] coordinates {(2.1,2.016437) (2.2,1.99955) (2.3,1.984515) (2.4,1.971193) (2.5,1.9593693) (2.6,1.948855) (2.7,1.9394840) (2.8,1.931099) (2.9,1.923578) (3.0,1.916802) (3.2,1.905132) (3.4,1.895497) (3.6,1.88745) (3.8,1.880665) (4.0,1.874885) (4.5,1.863721) (5.0,1.855775) (6.0,1.845481) (7.0,1.839308) (8.0,1.835316) (9.0,1.832585) (10.0,1.8306345) (11.0,1.829193) (12.0,1.828098) (13.0,1.8272465) (14.0,1.8265714) (15.0,1.8260268) (16.0,1.825581) (17.0,1.8252119) (18.0,1.8249025) (19.0,1.8246409) (20.0,1.824417) (21,1.824225) (22,1.8240586) (23,1.8239134) (24,1.82378576) (25,1.8236734) (26,1.823574) (27,1.823485) (28,1.823405) (29,1.8233336) (30.0,1.823269) (31,1.823211) (32,1.823158) (33,1.82311) (34,1.823066) (35,1.8230256) (36,1.8229884) (37,1.8229545) (38,1.8229234) (39,1.822894) (40,1.8228676)};

\addplot[domain=1:100, thick, dashed, ForestGreen] {1.8233};
\node[above left, text=ForestGreen] at (axis cs:35,1.8233) {Flat ESdS model};
\end{axis}
\end{tikzpicture}
\caption{Evolution of galaxy cluster mass $M$ with current cosmic scale factor $a_0$ in hyperbolic Einstein-Straus--de Sitter model. Cosmological constant $\Lambda$, angles $\alpha$ and $\alpha^{\prime}$ are fixed at their central values.}\label{Mvsa}
\end{figure}
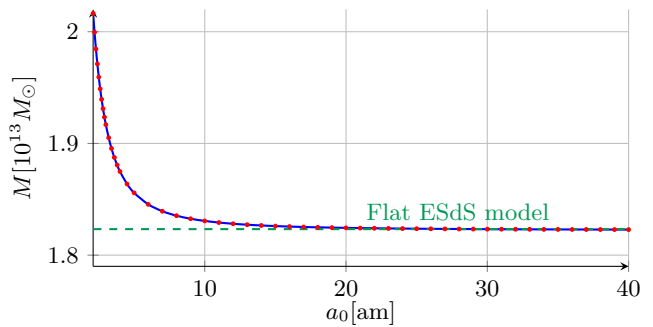

\begin{figure}[htbp]
\centering
\begin{tikzpicture}
\begin{axis} [axis lines=middle, width=8.7cm, height=5cm, xlabel={$a_0$[am]}, ylabel={$-\varphi_\text{S}[^{\prime\prime}]$}, xmin=2.1, xmax=40, ymin=9.94, ymax=10.1, grid=both, xlabel style={at={(axis description cs:0.5,-0.1)}, anchor=north}, ylabel style={at={(axis description cs:-0.09,0.5)}, anchor=south,rotate=90}]

\addplot[smooth, thick, blue] coordinates {(2.1,10.0961) (2.2,10.0913) (2.3,10.0854) (2.4,10.0792) (2.5,10.0731) (2.6,10.0673) (2.7,10.0617) (2.8,10.0565) (2.9,10.0517) (3.0,10.0472) (3.2,10.0392) (3.4,10.0322) (3.6,10.0263) (3.8,10.0212) (4.0,10.0167) (4.5,10.0079) (5.0,10.0015) (6.0,9.99297) (7.0,9.98776) (8.0,9.98435) (9.0,9.982) (10.0,9.98031) (11.0,9.97906) (12.0,9.97811) (13.0,9.97737) (14.0,9.97678) (15.0,9.9763) (16.0,9.97591) (17.0,9.97559) (18.0,9.97532) (19.0,9.97509) (20.0,9.97489) (21,9.97472) (22,9.97457) (23,9.97445) (24,9.97433) (25,9.97424) (26,9.97415) (27,9.97407) (28,9.974) (29,9.97394) (30.0,9.97388) (31,9.97383) (32,9.97378) (33,9.97374) (34,9.9737) (35,9.97367) (36,9.97363) (37,9.9736) (38,9.97358) (39,9.97355) (40,9.97353)};

\addplot[only marks, mark=*, red, mark size=0.7] coordinates {(2.1,10.0961) (2.2,10.0913) (2.3,10.0854) (2.4,10.0792) (2.5,10.0731) (2.6,10.0673) (2.7,10.0617) (2.8,10.0565) (2.9,10.0517) (3.0,10.0472) (3.2,10.0392) (3.4,10.0322) (3.6,10.0263) (3.8,10.0212) (4.0,10.0167) (4.5,10.0079) (5.0,10.0015) (6.0,9.99297) (7.0,9.98776) (8.0,9.98435) (9.0,9.982) (10.0,9.98031) (11.0,9.97906) (12.0,9.97811) (13.0,9.97737) (14.0,9.97678) (15.0,9.9763) (16.0,9.97591) (17.0,9.97559) (18.0,9.97532) (19.0,9.97509) (20.0,9.97489) (21,9.97472) (22,9.97457) (23,9.97445) (24,9.97433) (25,9.97424) (26,9.97415) (27,9.97407) (28,9.974) (29,9.97394) (30.0,9.97388) (31,9.97383) (32,9.97378) (33,9.97374) (34,9.9737) (35,9.97367) (36,9.97363) (37,9.9736) (38,9.97358) (39,9.97355) (40,9.97353)};

\addplot[domain=1:100, thick, dashed, ForestGreen] {9.9737};
\node[above left, text=ForestGreen] at (axis cs:35,9.9737) {Flat ESdS model};

\end{axis}
\end{tikzpicture}
\caption{Evolution of deflection angle $-\varphi_\text{S}$ with current cosmic scale factor $a_0$ in hyperbolic Einstein-Straus--de Sitter model. Cosmological constant $\Lambda$, angles $\alpha$ and $\alpha^{\prime}$ are fixed at their central values.}\label{defvsa}
\end{figure}

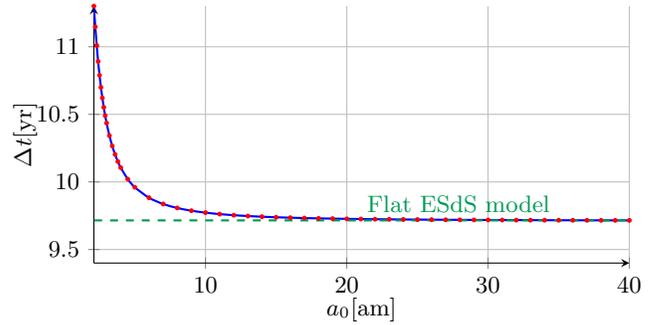
\begin{figure}[htbp]
\centering
\begin{tikzpicture}
\begin{axis} [axis lines=middle, width=8.7cm, height=5cm, xlabel={$a_0$[am]}, ylabel={$\Delta t[\text{yr}]$}, xmin=2.1, xmax=40, ymin=9.4, ymax=11.301, grid=both, xlabel style={at={(axis description cs:0.5,-0.1)}, anchor=north}, ylabel style={at={(axis description cs:-0.09,0.5)}, anchor=south,rotate=90}]

 \addplot[smooth, thick, blue] coordinates {(2.1,11.301) (2.2,11.1473) (2.3,11.01) (2.4,10.8918) (2.5,10.7893) (2.6,10.6999) (2.7,10.6215) (2.8,10.5522) (2.9,10.4907) (3.0,10.4359) (3.2,10.3425) (3.4,10.2663) (3.6,10.2034) (3.8,10.1507) (4.0,10.1061) (4.5,10.0207) (5.0,9.96047) (6.0,9.88309) (7.0,9.83703) (8.0,9.80736) (9.0,9.78712) (10.0,9.7727) (11.0,9.76206) (12.0,9.75398) (13.0,9.7477) (14.0,9.74273) (15.0,9.73872) (16.0,9.73544) (17.0,9.73272) (18.0,9.73044) (19.0,9.72852) (20.0,9.72688) (21,9.72546) (22,9.72424) (23,9.72317) (24,9.72224) (25,9.72141) (26,9.72068) (27,9.72002) (28,9.71944) (29,9.71891) (30.0,9.71844) (31,9.71801) (32,9.71762) (33,9.71727) (34,9.71695) (35,9.71665) (36,9.71638) (37,9.71613) (38,9.7159) (39,9.71569) (40,9.71549)};
 
 \addplot[only marks, mark=*, red, mark size=0.7] coordinates {(2.1,11.301) (2.2,11.1473) (2.3,11.01) (2.4,10.8918) (2.5,10.7893) (2.6,10.6999) (2.7,10.6215) (2.8,10.5522) (2.9,10.4907) (3.0,10.4359) (3.2,10.3425) (3.4,10.2663) (3.6,10.2034) (3.8,10.1507) (4.0,10.1061) (4.5,10.0207) (5.0,9.96047) (6.0,9.88309) (7.0,9.83703) (8.0,9.80736) (9.0,9.78712) (10.0,9.7727) (11.0,9.76206) (12.0,9.75398) (13.0,9.7477) (14.0,9.74273) (15.0,9.73872) (16.0,9.73544) (17.0,9.73272) (18.0,9.73044) (19.0,9.72852) (20.0,9.72688) (21,9.72546) (22,9.72424) (23,9.72317) (24,9.72224) (25,9.72141) (26,9.72068) (27,9.72002) (28,9.71944) (29,9.71891) (30.0,9.71844) (31,9.71801) (32,9.71762) (33,9.71727) (34,9.71695) (35,9.71665) (36,9.71638) (37,9.71613) (38,9.7159) (39,9.71569) (40,9.71549)};

\addplot[domain=1:100, thick, dashed, ForestGreen] {9.7155};
\node[above left, text=ForestGreen] at (axis cs:35,9.7155) {Flat ESdS model};

\end{axis}
\end{tikzpicture}
\caption{Evolution of time delay $\Delta t$ with current cosmic scale factor $a_0$ in hyperbolic Einstein-Straus--de Sitter model. Cosmological constant $\Lambda$, angles $\alpha$ and $\alpha^{\prime}$ are fixed at their central values.}\label{tdvsa}
\end{figure}

\begin{figure}[htbp]
\centering
\begin{tikzpicture}
\begin{axis} [axis lines=middle, x dir=reverse, width=8.7cm, height=5cm, xlabel={$|\Omega_{k0}|$}, ylabel={$M[10^{13} M_{\odot }]$}, xmin=2.1, xmax=40, ymin=1.79, ymax=2.02, grid=both, xlabel style={at={(axis description cs:0.5,-0.1)}, anchor=north}, ylabel style={at={(axis description cs:1.1,0.5)}, anchor=south,rotate=90},xticklabel={\pgfmathparse{(\tick)^(-2)}\pgfmathprintnumber[fixed,precision=4]{\pgfmathresult}}]

 \addplot[smooth, thick, blue] coordinates {(2.1,2.016437) (2.2,1.99955) (2.3,1.984515) (2.4,1.971193) (2.5,1.9593693) (2.6,1.948855) (2.7,1.9394840) (2.8,1.931099) (2.9,1.923578) (3.0,1.916802) (3.2,1.905132) (3.4,1.895497) (3.6,1.88745) (3.8,1.880665) (4.0,1.874885) (4.5,1.863721) (5.0,1.855775) (6.0,1.845481) (7.0,1.839308) (8.0,1.835316) (9.0,1.832585) (10.0,1.8306345) (11.0,1.829193) (12.0,1.828098) (13.0,1.8272465) (14.0,1.8265714) (15.0,1.8260268) (16.0,1.825581) (17.0,1.8252119) (18.0,1.8249025) (19.0,1.8246409) (20.0,1.824417) (21,1.824225) (22,1.8240586) (23,1.8239134) (24,1.82378576) (25,1.8236734) (26,1.823574) (27,1.823485) (28,1.823405) (29,1.8233336) (30.0,1.823269) (31,1.823211) (32,1.823158) (33,1.82311) (34,1.823066) (35,1.8230256) (36,1.8229884) (37,1.8229545) (38,1.8229234) (39,1.822894) (40,1.8228676)};
 
\addplot[only marks, mark=*, red, mark size=0.7] coordinates {(2.1,2.016437) (2.2,1.99955) (2.3,1.984515) (2.4,1.971193) (2.5,1.9593693) (2.6,1.948855) (2.7,1.9394840) (2.8,1.931099) (2.9,1.923578) (3.0,1.916802) (3.2,1.905132) (3.4,1.895497) (3.6,1.88745) (3.8,1.880665) (4.0,1.874885) (4.5,1.863721) (5.0,1.855775) (6.0,1.845481) (7.0,1.839308) (8.0,1.835316) (9.0,1.832585) (10.0,1.8306345) (11.0,1.829193) (12.0,1.828098) (13.0,1.8272465) (14.0,1.8265714) (15.0,1.8260268) (16.0,1.825581) (17.0,1.8252119) (18.0,1.8249025) (19.0,1.8246409) (20.0,1.824417) (21,1.824225) (22,1.8240586) (23,1.8239134) (24,1.82378576) (25,1.8236734) (26,1.823574) (27,1.823485) (28,1.823405) (29,1.8233336) (30.0,1.823269) (31,1.823211) (32,1.823158) (33,1.82311) (34,1.823066) (35,1.8230256) (36,1.8229884) (37,1.8229545) (38,1.8229234) (39,1.822894) (40,1.8228676)};

\addplot[domain=1:100, thick, dashed, ForestGreen] {1.8233};
\node[above right, text=ForestGreen] at (axis cs:35,1.8233) {Flat ESdS model};
\end{axis}
\end{tikzpicture}
\caption{Evolution of galaxy cluster mass $M$ with current curvature density $\Omega_{k0}$ in hyperbolic Einstein-Straus--de Sitter model. Cosmological constant $\Lambda$, angles $\alpha$ and $\alpha^{\prime}$ are fixed at their central values.}\label{MvsOmega}
\end{figure}
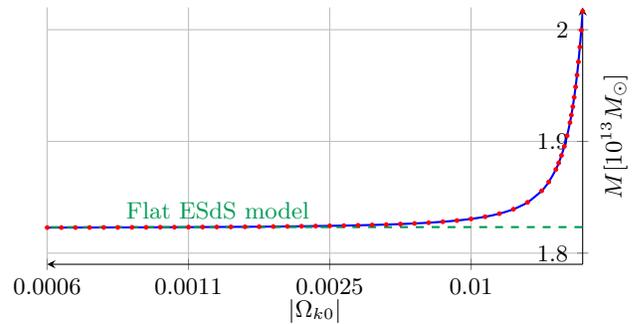

\begin{figure}[htbp]
\centering
\begin{tikzpicture}
\begin{axis} [axis lines=middle, x dir=reverse, width=8.7cm, height=5cm, xlabel={$|\Omega_{k0}|$}, ylabel={$-\varphi_\text{S}[^{\prime\prime}]$}, xmin=2.1, xmax=40, ymin=9.94, ymax=10.1, grid=both, xlabel style={at={(axis description cs:0.5,-0.1)}, anchor=north}, ylabel style={at={(axis description cs:1.1,0.5)}, anchor=south,rotate=90},xticklabel={\pgfmathparse{(\tick)^(-2)}\pgfmathprintnumber[fixed,precision=4]{\pgfmathresult}}]

\addplot[smooth, thick, blue] coordinates {(2.1,10.0961) (2.2,10.0913) (2.3,10.0854) (2.4,10.0792) (2.5,10.0731) (2.6,10.0673) (2.7,10.0617) (2.8,10.0565) (2.9,10.0517) (3.0,10.0472) (3.2,10.0392) (3.4,10.0322) (3.6,10.0263) (3.8,10.0212) (4.0,10.0167) (4.5,10.0079) (5.0,10.0015) (6.0,9.99297) (7.0,9.98776) (8.0,9.98435) (9.0,9.982) (10.0,9.98031) (11.0,9.97906) (12.0,9.97811) (13.0,9.97737) (14.0,9.97678) (15.0,9.9763) (16.0,9.97591) (17.0,9.97559) (18.0,9.97532) (19.0,9.97509) (20.0,9.97489) (21,9.97472) (22,9.97457) (23,9.97445) (24,9.97433) (25,9.97424) (26,9.97415) (27,9.97407) (28,9.974) (29,9.97394) (30.0,9.97388) (31,9.97383) (32,9.97378) (33,9.97374) (34,9.9737) (35,9.97367) (36,9.97363) (37,9.9736) (38,9.97358) (39,9.97355) (40,9.97353)};

\addplot[only marks, mark=*, red, mark size=0.7] coordinates {(2.1,10.0961) (2.2,10.0913) (2.3,10.0854) (2.4,10.0792) (2.5,10.0731) (2.6,10.0673) (2.7,10.0617) (2.8,10.0565) (2.9,10.0517) (3.0,10.0472) (3.2,10.0392) (3.4,10.0322) (3.6,10.0263) (3.8,10.0212) (4.0,10.0167) (4.5,10.0079) (5.0,10.0015) (6.0,9.99297) (7.0,9.98776) (8.0,9.98435) (9.0,9.982) (10.0,9.98031) (11.0,9.97906) (12.0,9.97811) (13.0,9.97737) (14.0,9.97678) (15.0,9.9763) (16.0,9.97591) (17.0,9.97559) (18.0,9.97532) (19.0,9.97509) (20.0,9.97489) (21,9.97472) (22,9.97457) (23,9.97445) (24,9.97433) (25,9.97424) (26,9.97415) (27,9.97407) (28,9.974) (29,9.97394) (30.0,9.97388) (31,9.97383) (32,9.97378) (33,9.97374) (34,9.9737) (35,9.97367) (36,9.97363) (37,9.9736) (38,9.97358) (39,9.97355) (40,9.97353)};

\addplot[domain=1:100, thick, dashed, ForestGreen] {9.9737};
\node[above right, text=ForestGreen] at (axis cs:35,9.9737) {Flat ESdS model};

\end{axis}
\end{tikzpicture}
\caption{Evolution of deflection angle $-\varphi_\text{S}$ with current curvature density $\Omega_{k0}$ in hyperbolic Einstein-Straus--de Sitter model. Cosmological constant $\Lambda$, angles $\alpha$ and $\alpha^{\prime}$ are fixed at their central values.}\label{defvsOmega}
\end{figure}

\begin{figure}[htbp]
\centering
\begin{tikzpicture}
\begin{axis} [axis lines=middle, x dir=reverse, width=8.7cm, height=5cm, xlabel={$|\Omega_{k0}|$}, ylabel={$\Delta t[\text{yr}]$}, xmin=2.1, xmax=40, ymin=9.4, ymax=11.301, grid=both, xlabel style={at={(axis description cs:0.5,-0.1)}, anchor=north}, ylabel style={at={(axis description cs:1.1,0.5)}, anchor=south,rotate=90},xticklabel={\pgfmathparse{(\tick)^(-2)}\pgfmathprintnumber[fixed,precision=4]{\pgfmathresult}}]

 \addplot[smooth, thick, blue] coordinates {(2.1,11.301) (2.2,11.1473) (2.3,11.01) (2.4,10.8918) (2.5,10.7893) (2.6,10.6999) (2.7,10.6215) (2.8,10.5522) (2.9,10.4907) (3.0,10.4359) (3.2,10.3425) (3.4,10.2663) (3.6,10.2034) (3.8,10.1507) (4.0,10.1061) (4.5,10.0207) (5.0,9.96047) (6.0,9.88309) (7.0,9.83703) (8.0,9.80736) (9.0,9.78712) (10.0,9.7727) (11.0,9.76206) (12.0,9.75398) (13.0,9.7477) (14.0,9.74273) (15.0,9.73872) (16.0,9.73544) (17.0,9.73272) (18.0,9.73044) (19.0,9.72852) (20.0,9.72688) (21,9.72546) (22,9.72424) (23,9.72317) (24,9.72224) (25,9.72141) (26,9.72068) (27,9.72002) (28,9.71944) (29,9.71891) (30.0,9.71844) (31,9.71801) (32,9.71762) (33,9.71727) (34,9.71695) (35,9.71665) (36,9.71638) (37,9.71613) (38,9.7159) (39,9.71569) (40,9.71549)};
 
 \addplot[only marks, mark=*, red, mark size=0.7] coordinates {(2.1,11.301) (2.2,11.1473) (2.3,11.01) (2.4,10.8918) (2.5,10.7893) (2.6,10.6999) (2.7,10.6215) (2.8,10.5522) (2.9,10.4907) (3.0,10.4359) (3.2,10.3425) (3.4,10.2663) (3.6,10.2034) (3.8,10.1507) (4.0,10.1061) (4.5,10.0207) (5.0,9.96047) (6.0,9.88309) (7.0,9.83703) (8.0,9.80736) (9.0,9.78712) (10.0,9.7727) (11.0,9.76206) (12.0,9.75398) (13.0,9.7477) (14.0,9.74273) (15.0,9.73872) (16.0,9.73544) (17.0,9.73272) (18.0,9.73044) (19.0,9.72852) (20.0,9.72688) (21,9.72546) (22,9.72424) (23,9.72317) (24,9.72224) (25,9.72141) (26,9.72068) (27,9.72002) (28,9.71944) (29,9.71891) (30.0,9.71844) (31,9.71801) (32,9.71762) (33,9.71727) (34,9.71695) (35,9.71665) (36,9.71638) (37,9.71613) (38,9.7159) (39,9.71569) (40,9.71549)};

\addplot[domain=1:100, thick, dashed, ForestGreen] {9.7155};
\node[above right, text=ForestGreen] at (axis cs:35,9.7155) {Flat ESdS model};

\end{axis}
\end{tikzpicture}
\caption{Evolution of time delay $\Delta t$ with current curvature density $\Omega_{k0}$ in hyperbolic Einstein-Straus--de Sitter model. Cosmological constant $\Lambda$, angles $\alpha$ and $\alpha^{\prime}$ are fixed at their central values.}\label{tdvsOmega}
\end{figure}
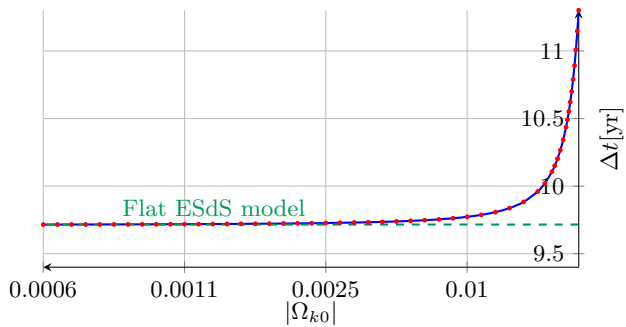

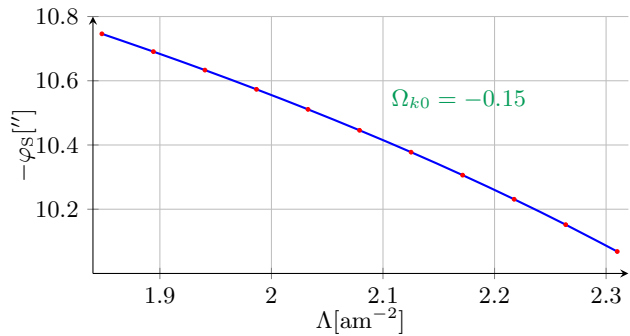
\begin{figure}[htbp]
\centering
\begin{tikzpicture}
\begin{axis} [axis lines=middle, width=8.7cm, height=5cm, xlabel={$\Lambda [\text{am}^{-2}]$}, ylabel={$-\varphi_\text{S}$[$^{\prime\prime}$]}, xmin=1.84, xmax=2.32, ymin=10.00, ymax=10.8, grid=both, xlabel style={at={(axis description cs:0.5,-0.1)}, anchor=north}, ylabel style={at={(axis description cs:-0.09,0.5)}, anchor=south,rotate=90}]

\addplot[smooth, thick, blue] coordinates {(1.848,10.7461) (1.8942,10.6907) (1.9404,10.6332) (1.9866,10.5733) (2.0328,10.5109) (2.079,10.4457) (2.1252,10.3776) (2.1714,10.3061) (2.2176,10.231) (2.2638,10.1519) (2.31,10.0683)};

\addplot[only marks, mark=*, red, mark size=0.7] coordinates {(1.848,10.7461) (1.8942,10.6907) (1.9404,10.6332) (1.9866,10.5733) (2.0328,10.5109) (2.079,10.4457) (2.1252,10.3776) (2.1714,10.3061) (2.2176,10.231) (2.2638,10.1519) (2.31,10.0683)};

\node[below right, text=ForestGreen] at (axis cs:2.1,10.6) {$\Omega_{k0}=-0.15$};

\end{axis}
\end{tikzpicture}
\caption{Evolution of deflection angle $-\varphi_\text{S}$ as function of cosmological constant $\Lambda$ within the range $[1.848\,\text{am}^{-2},2.31\,\text{am}^{-2}]$ for negative curvature density $\Omega_{k0}=-0.15$ in hyperbolic Einstein-Straus--de Sitter model with angles $\alpha$ and $\alpha^{\prime}$ fixed at their central values.}\label{defvsLambda}
\end{figure}

To quantify the influence of $\Lambda$ on light bending, Table~\ref{tab1} has been made to present the adjusted mass and the deflection angle $\varphi_\text{S}$ $(=\varphi_\text{S}^\prime)$ for the maximal $+$, central $\pm0$, and minimal $-$ values of $\Lambda$ as well as the angles $\alpha$ and $\alpha^\prime$, employing a fixed current scale factor at $a_0=5\,\text{am}$, corresponding to $\Omega_{k0}=-0.04$). For each configuration, the time delay of the photon $\alpha$ relative to the photon $\alpha^\prime$ is computed. Note that neither $\Lambda$ nor $a_0$ adjustments reconcile the $\varphi_\text{S}$ angles.

The results presented in Table~\ref{tab1} can be succinctly  summarized by the following parameter ranges: $1.48\times 10^{13}M_{\odot }\leq M\leq 2.25\times 10^{13}M_{\odot }$, $
6.31^{\prime \prime }\leq -\varphi_\text{S}\leq 13.79^{\prime \prime }$, and $6.75\,\text{yr}\leq \Delta t\leq 13.10\,\text{yr}$. The galaxy cluster mass is underestimated compared to observations \cite{Inada,Oguri}, due to the spherical lens assumption. The predicted time delays are comparable with the observational lower bound $\Delta t_\text{CD}> 7.7\,\text{yr}$ from Fohlmeister \cite{Fohl} as well as the upper bound $\Delta t_\text{CD}\lesssim 10.14\,\text{yr}$ from Kawano and Oguri \cite{Kawano}, who employed a nonspherical lens model.

An analysis of  Table~\ref{tab1} reveals several important quantitative effects of varying the cosmological constant on gravitational lensing by a galaxy cluster. A $20\%$ increase in the cosmological constant $\Lambda$ reduces the deflection angle $\varphi_\text{S}$ by approximately $9.02\%$, while the galaxy cluster mass changes by only $0.28\%$. For comparison, this reduction slightly exceeds the effect observed for $\Omega_{k0}= 0.04$ in the positively curved case \cite{Guen2}, where the deflection angle is diminished by $8.03\%$. We therefore conclude that a positive $\Lambda$ weakens light deflection in agreement with the statement by Rindler and Ishak \cite{Rind}.

Table~\ref{tab2} shows how the mass, deflection angle, and time delay vary depending on different values of $a_0$ over the range $[2.1\,\text{am},40\,\text{am}]$, while holding $\Lambda$, $\alpha$ and $\alpha^\prime$ fixed at their central values. To better visualize this dependence, we present interpolated plots in Figs.~\ref{Mvsa}, \ref{defvsa}, and \ref{tdvsa} for the datasets $(M,a_0)$, $(-\varphi_\text{S},a_0)$, and $(\Delta t,a_0)$, respectively, as well as their corresponding plots as functions of the modulus of the current curvature density $|\Omega_{k0}|$ in Figs.~\ref{MvsOmega}, \ref{defvsOmega}, and \ref{tdvsOmega}. Contrary to the flat ESdS model, where the lensing observables remain the same regardless of the current scale factor value, the results obtained in the hyperbolic ESdS model indicate that the mass, deflection angle, and time delay increase as $a_0$ becomes smaller than a limiting value close to $20 \,\text{am}$ (corresponding to a current curvature density $\Omega_{k0}\simeq -0.0025$), while above this limit, the results become independent of $a_0$ and compatible with the flat case \cite{Guen1}. The same applies for all possible values of $\Lambda$, $\alpha$ and $\alpha^\prime$ within their error bars. Particularly, for a negative curvature density $\Omega_{k0}=-0.15$, the light bending increases slightly by about $0.94\%$, whereas time delay increases significantly by about $10.17\%$. Comparing to the positively curved case \cite{Guen2}, with a positive curvature density $\Omega_{k0}=0.15$, the light bending and time delay decrease by about $0.66\%$ and $8.91\%$ respectively, which are slightly less than the effect produced in the negatively curved case. It is worth noting that even with this relatively large negative curvature, which tends to increase lensing observables, varying the value of the cosmological constant within its error margin results in a reduction of the deflection angle by approximately $6.72\%$, as illustrated in Fig.~\ref{defvsLambda}. This highlights that the $\Lambda$ effect on light bending does not vanish even when the Universe has substantial negative curvature; rather, it still acts to reduce light bending in the same direction as in flat or positively curved universes. Regarding the observed asymmetry—where negative spatial curvature significantly affects the time delay but only slightly alters the deflection angle—it has a physical explanation. The deflection angle is dominated by the local bending of the photon trajectory near the lens, where the gravitational potential of the mass distribution is the main contributor. Negative curvature primarily affects the large-scale geometry of the FLRW background, which induces only small corrections to this local bending. In contrast, the time delay has both a geometrical component (path-length difference between the photon trajectories) and a Shapiro component. Negative curvature stretches the background spacetime, effectively increasing the comoving distance that photons must travel. This stretching amplifies the geometrical part of the time delay more strongly than the localized deflection angle. In other words, while the lensing deflection is local and hence less sensitive to the global curvature, the time delay accumulates curvature effects along the entire light path and thus responds more strongly.

\FloatBarrier
\section{\label{sec5}Conclusion}
This work extends gravitational lensing analysis to the negatively curved Einstein-Straus--de Sitter model, providing distinct predictions for light deflection and time delay compared to the flat and positively curved models. The model uses a Swiss cheese approach, whereby a static SdS vacuole is matched to an expanding FLRW universe with negative spatial curvature $k=-1$. 

Throughout the computation of light deflection and time delay, we employed a hybrid analytical-numerical integration of differential equations in each region to address the nonflat case.

An application to the lensed quasar SDSS J1004+4112 permits one to quantify the subtle interplay between the spatial curvature and the cosmological constant in gravitational lensing phenomena. Negative spatial curvature ($k=-1$) enhances effective comoving distances, thereby increasing lensing observables, whereas a positive $\Lambda$ opposes this enhancement by introducing a repulsive contribution that reduces light deflection, as established in previous works. This reduction occurs because the repulsive effect of $\Lambda$ counteracts the spacetime curvature induced by mass. The study, based on varying the current scale factor $a_0$ (or equivalently the current curvature density), reveals a critical threshold of $a_0\simeq2.6\times10^{27}\,\text{m}$, below which lensing observables increase markedly, linked to $\Omega_{k0}\lesssim-0.0025$. Beyond this threshold, results match the flat ESdS model. In addition, the study reveals that the effect of $\Lambda$ on lensing does persist even in scenarios dominated by significant negative curvature and should be included in precise lensing analyses. Specifically, for a negative curvature density $\Omega_{k0}=-0.15$, the light bending increases slightly by about $0.94\%$, whereas time delay increases significantly by about $10.17\%$. This disparity can be explained by the fact that the deflection angle is dominated by local bending near the lens and is only weakly sensitive to the large-scale background curvature, whereas the time delay contains a geometric component that accumulates along the entire photon path and is therefore more strongly impacted by the global curvature. This behavior provides a unique signature for the negatively curved universe, distinct from flat and positively curved cases. Such a sensitivity to the curvature parameter means that precise lensing observations could, in principle, help discriminate between open, flat, and closed universes. In hyperbolic space, light rays that are initially close together diverge more rapidly than in flat ($k=0$) or positively curved ($k=+1$) spaces. This property leads to longer comoving distances for a given redshift and larger apparent angular separations at the observer for the same lensing configuration, thereby amplifying the cumulative effects of gravitational bending—both deflection angles and time delays—relative to flat or positively curved geometries. These deviations, though subtle, could provide a way to constrain the value of $\Omega_{k0}$ by observing differences in lensing observables from what is expected in a flat universe. This sensitivity to the curvature parameter suggests that, in principle, future high-precision lensing observations, when combined with independent cosmological probes and realistic lens modeling, could contribute to discriminating between open, flat, and closed universes. However, the predicted variations in lensing observables for realistic $|\Omega_{k0}|$ are typically at the subpercent to few-percent level, requiring extremely accurate measurements and detailed modeling of lens mass distributions to detect. In practice, such tests would need to be carried out statistically over large samples of strong-lensing systems and combined with other probes, such as CMB and BAO constraints, to break degeneracies with other cosmological parameters.

Nevertheless, our analysis is subject to certain limitations. The assumption of a spherically symmetric lens and the use of the Swiss-cheese model, while physically motivated, may underestimate the complexity of real astrophysical systems. Future work should extend the present approach to include more realistic mass distributions, and to consider additional lensed systems with well-constrained observational data.

Another promising direction for future work is to incorporate the interior SdS solution, which describes the spacetime inside a spherically incompressible mass with a cosmological constant. Unlike the exterior case, the interior solution introduces a direct dependence of light deflection on $\Lambda$, leading to subtle corrections in lensing observables, as shown by Schücker \cite{Schu3}. Incorporating this effect would yield a more complete and accurate description of strong lensing by massive objects, and could help refine cosmological constraints through high-precision lensing measurements \cite{Guen4,Guen5}.

Finally, our analysis of light deflection and time delay in an open, negatively curved Einstein-Straus–de Sitter universe complements a broader effort to probe classical and quantum gravitational effects on lensing observables. Seminal studies by Virbhadra and collaborators have shown that strong lensing near black holes produces distinctive image configurations and time‑delay signatures highly sensitive to spacetime geometry \cite{Virbhadra2000,Virbhadra2002,VirbhadraEllis2009}. More recent work has explored quantum corrections to general relativity in strong field regimes, predicting subtle shifts in the Kerr time delay \cite{Battista2017}, while investigations of loop quantum gravity-inspired black holes with holonomy corrections, global monopole charges, or topological defects have revealed nontrivial modifications of deflection angles and time delays in photon‑sphere and high‑precision lensing configurations \cite{Soares2023a,Soares2023b,Soares2023c,Soares2023d}. Finally, topologically charged Eddington‑inspired Born–Infeld wormhole models \cite{Soares2023c} further illustrate how nontrivial global structure can affect lensing profiles, underscoring the need for flexible modeling frameworks like the one employed here to capture geometric effects in nonasymptotically flat settings.

\begin{acknowledgments}
The author would like to thank the anonymous referee for valuable comments and suggestions that helped improve the quality and clarity of this work.
\end{acknowledgments}

\bibliography{references}

\begin{thebibliography}{49}%
\makeatletter
\providecommand \@ifxundefined [1]{%
 \@ifx{#1\undefined}
}%
\providecommand \@ifnum [1]{%
 \ifnum #1\expandafter \@firstoftwo
 \else \expandafter \@secondoftwo
 \fi
}%
\providecommand \@ifx [1]{%
 \ifx #1\expandafter \@firstoftwo
 \else \expandafter \@secondoftwo
 \fi
}%
\providecommand \natexlab [1]{#1}%
\providecommand \enquote  [1]{``#1''}%
\providecommand \bibnamefont  [1]{#1}%
\providecommand \bibfnamefont [1]{#1}%
\providecommand \citenamefont [1]{#1}%
\providecommand \href@noop [0]{\@secondoftwo}%
\providecommand \href [0]{\begingroup \@sanitize@url \@href}%
\providecommand \@href[1]{\@@startlink{#1}\@@href}%
\providecommand \@@href[1]{\endgroup#1\@@endlink}%
\providecommand \@sanitize@url [0]{\catcode `\\12\catcode `\$12\catcode `\&12\catcode `\#12\catcode `\^12\catcode `\_12\catcode `\%12\relax}%
\providecommand \@@startlink[1]{}%
\providecommand \@@endlink[0]{}%
\providecommand \url  [0]{\begingroup\@sanitize@url \@url }%
\providecommand \@url [1]{\endgroup\@href {#1}{\urlprefix }}%
\providecommand \urlprefix  [0]{URL }%
\providecommand \Eprint [0]{\href }%
\providecommand \doibase [0]{https://doi.org/}%
\providecommand \selectlanguage [0]{\@gobble}%
\providecommand \bibinfo  [0]{\@secondoftwo}%
\providecommand \bibfield  [0]{\@secondoftwo}%
\providecommand \translation [1]{[#1]}%
\providecommand \BibitemOpen [0]{}%
\providecommand \bibitemStop [0]{}%
\providecommand \bibitemNoStop [0]{.\EOS\space}%
\providecommand \EOS [0]{\spacefactor3000\relax}%
\providecommand \BibitemShut  [1]{\csname bibitem#1\endcsname}%
\let\auto@bib@innerbib\@empty
\bibitem [{\citenamefont {Aghanim}\ \emph {et~al.}(2020)\citenamefont {Aghanim} \emph {et~al.}}]{Planck2020}%
  \BibitemOpen
  \bibfield  {author} {\bibinfo {author} {\bibfnamefont {N.}~\bibnamefont {Aghanim}} \emph {et~al.} (\bibinfo {collaboration} {Planck Collaboration}),\ }\bibfield  {title} {\bibinfo {title} {Planck 2018 results. {VI}. {Cosmological} parameters},\ }\href {https://doi.org/10.1051/0004-6361/201833910} {\bibfield  {journal} {\bibinfo  {journal} {Astron. Astrophys.}\ }\textbf {\bibinfo {volume} {641}},\ \bibinfo {pages} {A6} (\bibinfo {year} {2020})}\BibitemShut {NoStop}%
\bibitem [{\citenamefont {Valentino}\ \emph {et~al.}(2021)\citenamefont {Valentino}, \citenamefont {Melchiorri},\ and\ \citenamefont {Silk}}]{DiValentino2021}%
  \BibitemOpen
  \bibfield  {author} {\bibinfo {author} {\bibfnamefont {E.~D.}\ \bibnamefont {Valentino}}, \bibinfo {author} {\bibfnamefont {A.}~\bibnamefont {Melchiorri}},\ and\ \bibinfo {author} {\bibfnamefont {J.}~\bibnamefont {Silk}},\ }\bibfield  {title} {\bibinfo {title} {In the realm of the {Hubble} tension—a review of solutions},\ }\href {https://doi.org/10.1088/1361-6382/ac086d} {\bibfield  {journal} {\bibinfo  {journal} {Class. Quant. Grav.}\ }\textbf {\bibinfo {volume} {38}},\ \bibinfo {pages} {153001} (\bibinfo {year} {2021})}\BibitemShut {NoStop}%
\bibitem [{\citenamefont {Linde}(1982)}]{Linde1982}%
  \BibitemOpen
  \bibfield  {author} {\bibinfo {author} {\bibfnamefont {A.~D.}\ \bibnamefont {Linde}},\ }\bibfield  {title} {\bibinfo {title} {A new inflationary universe scenario: A possible solution of the horizon, flatness, homogeneity, isotropy and primordial monopole problems},\ }\href {https://doi.org/10.1016/0370-2693(82)91219-9} {\bibfield  {journal} {\bibinfo  {journal} {Phys. Lett. B}\ }\textbf {\bibinfo {volume} {108}},\ \bibinfo {pages} {389} (\bibinfo {year} {1982})}\BibitemShut {NoStop}%
\bibitem [{\citenamefont {Linde}(1986)}]{Linde1986}%
  \BibitemOpen
  \bibfield  {author} {\bibinfo {author} {\bibfnamefont {A.~D.}\ \bibnamefont {Linde}},\ }\bibfield  {title} {\bibinfo {title} {Eternally existing self-reproducing chaotic inflationary universe},\ }\href {https://doi.org/10.1016/0370-2693(86)90611-8} {\bibfield  {journal} {\bibinfo  {journal} {Phys. Lett. B}\ }\textbf {\bibinfo {volume} {175}},\ \bibinfo {pages} {395} (\bibinfo {year} {1986})}\BibitemShut {NoStop}%
\bibitem [{\citenamefont {Freivogel}\ \emph {et~al.}(2006)\citenamefont {Freivogel}, \citenamefont {Kleban}, \citenamefont {Martínez},\ and\ \citenamefont {Susskind}}]{Freivogel2006}%
  \BibitemOpen
  \bibfield  {author} {\bibinfo {author} {\bibfnamefont {B.}~\bibnamefont {Freivogel}}, \bibinfo {author} {\bibfnamefont {M.}~\bibnamefont {Kleban}}, \bibinfo {author} {\bibfnamefont {M.~R.}\ \bibnamefont {Martínez}},\ and\ \bibinfo {author} {\bibfnamefont {L.}~\bibnamefont {Susskind}},\ }\bibfield  {title} {\bibinfo {title} {Observational consequences of a landscape},\ }\href {https://doi.org/10.1088/1126-6708/2006/03/039} {\bibfield  {journal} {\bibinfo  {journal} {J. High Energy Phys.}\ }\textbf {\bibinfo {volume} {03}}\bibinfo  {number} { (2006)},\ \bibinfo {pages} {039}}\BibitemShut {NoStop}%
\bibitem [{\citenamefont {Guth}(1981)}]{Guth1981}%
  \BibitemOpen
\bibfield  {number} {  }\bibfield  {author} {\bibinfo {author} {\bibfnamefont {A.~H.}\ \bibnamefont {Guth}},\ }\bibfield  {title} {\bibinfo {title} {Inflationary universe: A possible solution to the horizon and flatness problems},\ }\href {https://doi.org/10.1103/PhysRevD.23.347} {\bibfield  {journal} {\bibinfo  {journal} {Phys. Rev. D}\ }\textbf {\bibinfo {volume} {23}},\ \bibinfo {pages} {347} (\bibinfo {year} {1981})}\BibitemShut {NoStop}%
\bibitem [{\citenamefont {Einstein}\ and\ \citenamefont {Straus}(1945)}]{ES1}%
  \BibitemOpen
  \bibfield  {author} {\bibinfo {author} {\bibfnamefont {A.}~\bibnamefont {Einstein}}\ and\ \bibinfo {author} {\bibfnamefont {E.~G.}\ \bibnamefont {Straus}},\ }\bibfield  {title} {\bibinfo {title} {The influence of the expansion of space on the gravitation fields surrounding the individual star},\ }\href {https://doi.org/10.1103/RevModPhys.17.120} {\bibfield  {journal} {\bibinfo  {journal} {Rev. Mod. Phys.}\ }\textbf {\bibinfo {volume} {17}},\ \bibinfo {pages} {120} (\bibinfo {year} {1945})}\BibitemShut {NoStop}%
\bibitem [{\citenamefont {Einstein}\ and\ \citenamefont {Straus}(1946)}]{ES2}%
  \BibitemOpen
  \bibfield  {author} {\bibinfo {author} {\bibfnamefont {A.}~\bibnamefont {Einstein}}\ and\ \bibinfo {author} {\bibfnamefont {E.~G.}\ \bibnamefont {Straus}},\ }\bibfield  {title} {\bibinfo {title} {Corrections and additional remarks to our paper: The influence of the expansion of space on the gravitation fields surrounding the individual stars},\ }\href {https://doi.org/10.1103/RevModPhys.18.148} {\bibfield  {journal} {\bibinfo  {journal} {Rev. Mod. Phys.}\ }\textbf {\bibinfo {volume} {18}},\ \bibinfo {pages} {148} (\bibinfo {year} {1946})}\BibitemShut {NoStop}%
\bibitem [{\citenamefont {Balbinot}\ \emph {et~al.}(1988)\citenamefont {Balbinot}, \citenamefont {Bergamini},\ and\ \citenamefont {Comastri}}]{Balbinot}%
  \BibitemOpen
  \bibfield  {author} {\bibinfo {author} {\bibfnamefont {R.}~\bibnamefont {Balbinot}}, \bibinfo {author} {\bibfnamefont {R.}~\bibnamefont {Bergamini}},\ and\ \bibinfo {author} {\bibfnamefont {A.}~\bibnamefont {Comastri}},\ }\bibfield  {title} {\bibinfo {title} {Solution of the einstein-strauss problem with a {$\Lambda$} term},\ }\href {https://doi.org/10.1103/PhysRevD.38.2415} {\bibfield  {journal} {\bibinfo  {journal} {Phys. Rev. D}\ }\textbf {\bibinfo {volume} {38}},\ \bibinfo {pages} {2415} (\bibinfo {year} {1988})}\BibitemShut {NoStop}%
\bibitem [{\citenamefont {Guenouche}(2024)}]{Guen3}%
  \BibitemOpen
  \bibfield  {author} {\bibinfo {author} {\bibfnamefont {M.}~\bibnamefont {Guenouche}},\ }\bibfield  {title} {\bibinfo {title} {Effect of the spatial curvature on light bending and time delay in a curved {Einstein-Straus--de Sitter} spacetime},\ }\href {https://doi.org/10.1103/PhysRevD.110.063508} {\bibfield  {journal} {\bibinfo  {journal} {Phys. Rev. D}\ }\textbf {\bibinfo {volume} {110}},\ \bibinfo {pages} {063508} (\bibinfo {year} {2024})}\BibitemShut {NoStop}%
\bibitem [{\citenamefont {Rindler}\ and\ \citenamefont {Ishak}(2007)}]{Rind}%
  \BibitemOpen
  \bibfield  {author} {\bibinfo {author} {\bibfnamefont {W.}~\bibnamefont {Rindler}}\ and\ \bibinfo {author} {\bibfnamefont {M.}~\bibnamefont {Ishak}},\ }\bibfield  {title} {\bibinfo {title} {Contribution of the cosmological constant to the relativistic bending of light revisited},\ }\href {https://doi.org/10.1103/PhysRevD.76.043006} {\bibfield  {journal} {\bibinfo  {journal} {Phys. Rev. D}\ }\textbf {\bibinfo {volume} {76}},\ \bibinfo {pages} {043006} (\bibinfo {year} {2007})}\BibitemShut {NoStop}%
\bibitem [{\citenamefont {Ishak}\ \emph {et~al.}(2008)\citenamefont {Ishak}, \citenamefont {Rindler}, \citenamefont {Dossett}, \citenamefont {Moldenhauer},\ and\ \citenamefont {Allison}}]{Ishak}%
  \BibitemOpen
  \bibfield  {author} {\bibinfo {author} {\bibfnamefont {M.}~\bibnamefont {Ishak}}, \bibinfo {author} {\bibfnamefont {W.}~\bibnamefont {Rindler}}, \bibinfo {author} {\bibfnamefont {J.}~\bibnamefont {Dossett}}, \bibinfo {author} {\bibfnamefont {J.}~\bibnamefont {Moldenhauer}},\ and\ \bibinfo {author} {\bibfnamefont {C.}~\bibnamefont {Allison}},\ }\bibfield  {title} {\bibinfo {title} {A new independent limit on the cosmological constant/dark energy from the relativistic bending of light by galaxies and clusters of galaxies},\ }\href {https://doi.org/10.1111/j.1365-2966.2008.13468.x} {\bibfield  {journal} {\bibinfo  {journal} {Mon. Not. R. Astron. Soc.}\ }\textbf {\bibinfo {volume} {388}},\ \bibinfo {pages} {1279} (\bibinfo {year} {2008})}\BibitemShut {NoStop}%
\bibitem [{\citenamefont {Schücker}(2009{\natexlab{a}})}]{Schu4}%
  \BibitemOpen
  \bibfield  {author} {\bibinfo {author} {\bibfnamefont {T.}~\bibnamefont {Schücker}},\ }\bibfield  {title} {\bibinfo {title} {Cosmological constant and lensing},\ }\href {https://doi.org/10.1007/s10714-008-0652-2} {\bibfield  {journal} {\bibinfo  {journal} {Gen. Relativ. Gravit.}\ }\textbf {\bibinfo {volume} {41}},\ \bibinfo {pages} {67} (\bibinfo {year} {2009}{\natexlab{a}})}\BibitemShut {NoStop}%
\bibitem [{\citenamefont {Schücker}(2008)}]{Schu6}%
  \BibitemOpen
  \bibfield  {author} {\bibinfo {author} {\bibfnamefont {T.}~\bibnamefont {Schücker}},\ }\href@noop {} {\bibinfo {title} {Strong lensing with positive cosmological constant}} (\bibinfo {year} {2008}),\ \Eprint {https://arxiv.org/abs/arXiv:0805.1630} {arXiv:0805.1630} \BibitemShut {NoStop}%
\bibitem [{\citenamefont {Sereno}(2009)}]{Sereno}%
  \BibitemOpen
  \bibfield  {author} {\bibinfo {author} {\bibfnamefont {M.}~\bibnamefont {Sereno}},\ }\bibfield  {title} {\bibinfo {title} {Role of {$\Lambda$} in the cosmological lens equation},\ }\href {https://doi.org/10.1103/PhysRevLett.102.021301} {\bibfield  {journal} {\bibinfo  {journal} {Phys. Rev. Lett.}\ }\textbf {\bibinfo {volume} {102}},\ \bibinfo {pages} {021301} (\bibinfo {year} {2009})}\BibitemShut {NoStop}%
\bibitem [{\citenamefont {Khriplovich}\ and\ \citenamefont {Pomeransky}(2008)}]{Khrip}%
  \BibitemOpen
  \bibfield  {author} {\bibinfo {author} {\bibfnamefont {I.~B.}\ \bibnamefont {Khriplovich}}\ and\ \bibinfo {author} {\bibfnamefont {A.~A.}\ \bibnamefont {Pomeransky}},\ }\bibfield  {title} {\bibinfo {title} {Does the cosmological term influence gravitational lensing?},\ }\href {https://doi.org/10.1142/S0218271808013832} {\bibfield  {journal} {\bibinfo  {journal} {Int. J. Mod. Phys. D}\ }\textbf {\bibinfo {volume} {17}},\ \bibinfo {pages} {2255} (\bibinfo {year} {2008})}\BibitemShut {NoStop}%
\bibitem [{\citenamefont {Park}(2008)}]{Park}%
  \BibitemOpen
  \bibfield  {author} {\bibinfo {author} {\bibfnamefont {M.}~\bibnamefont {Park}},\ }\bibfield  {title} {\bibinfo {title} {Rigorous approach to gravitational lensing},\ }\href {https://doi.org/10.1103/PhysRevD.78.023014} {\bibfield  {journal} {\bibinfo  {journal} {Phys. Rev. D}\ }\textbf {\bibinfo {volume} {78}},\ \bibinfo {pages} {023014} (\bibinfo {year} {2008})}\BibitemShut {NoStop}%
\bibitem [{\citenamefont {Simpson}\ \emph {et~al.}(2010)\citenamefont {Simpson}, \citenamefont {Peacock},\ and\ \citenamefont {Heavens}}]{Simpson}%
  \BibitemOpen
  \bibfield  {author} {\bibinfo {author} {\bibfnamefont {F.}~\bibnamefont {Simpson}}, \bibinfo {author} {\bibfnamefont {J.~A.}\ \bibnamefont {Peacock}},\ and\ \bibinfo {author} {\bibfnamefont {A.~F.}\ \bibnamefont {Heavens}},\ }\bibfield  {title} {\bibinfo {title} {On lensing by a cosmological constant},\ }\href {https://doi.org/10.1111/j.1365-2966.2009.16032.x} {\bibfield  {journal} {\bibinfo  {journal} {Mon. Not. R. Astron. Soc.}\ }\textbf {\bibinfo {volume} {402}},\ \bibinfo {pages} {2009} (\bibinfo {year} {2010})}\BibitemShut {NoStop}%
\bibitem [{\citenamefont {Kantowski}\ \emph {et~al.}(2010)\citenamefont {Kantowski}, \citenamefont {Chen},\ and\ \citenamefont {Dai}}]{Kant1}%
  \BibitemOpen
  \bibfield  {author} {\bibinfo {author} {\bibfnamefont {R.}~\bibnamefont {Kantowski}}, \bibinfo {author} {\bibfnamefont {B.}~\bibnamefont {Chen}},\ and\ \bibinfo {author} {\bibfnamefont {X.}~\bibnamefont {Dai}},\ }\bibfield  {title} {\bibinfo {title} {Gravitational lensing corrections in flat {$\Lambda$CDM} cosmology},\ }\href {https://doi.org/10.1088/0004-637X/718/2/913} {\bibfield  {journal} {\bibinfo  {journal} {Astrophys. J.}\ }\textbf {\bibinfo {volume} {718}},\ \bibinfo {pages} {913} (\bibinfo {year} {2010})}\BibitemShut {NoStop}%
\bibitem [{\citenamefont {Schücker}(2010{\natexlab{a}})}]{Schu5}%
  \BibitemOpen
  \bibfield  {author} {\bibinfo {author} {\bibfnamefont {T.}~\bibnamefont {Schücker}},\ }\href@noop {} {\bibinfo {title} {Lensing in the {Einstein-Straus} solution}} (\bibinfo {year} {2010}{\natexlab{a}}),\ \Eprint {https://arxiv.org/abs/arXiv:1006.3234} {arXiv:1006.3234} \BibitemShut {NoStop}%
\bibitem [{\citenamefont {Chen}\ \emph {et~al.}(2010)\citenamefont {Chen}, \citenamefont {Kantowski},\ and\ \citenamefont {Dai}}]{Chen}%
  \BibitemOpen
  \bibfield  {author} {\bibinfo {author} {\bibfnamefont {B.}~\bibnamefont {Chen}}, \bibinfo {author} {\bibfnamefont {R.}~\bibnamefont {Kantowski}},\ and\ \bibinfo {author} {\bibfnamefont {X.}~\bibnamefont {Dai}},\ }\bibfield  {title} {\bibinfo {title} {Gravitational lens equation for embedded lenses; magnification and ellipticity},\ }\href {https://doi.org/10.1103/PhysRevD.84.083004} {\bibfield  {journal} {\bibinfo  {journal} {Phys. Rev. D}\ }\textbf {\bibinfo {volume} {84}},\ \bibinfo {pages} {083004} (\bibinfo {year} {2010})}\BibitemShut {NoStop}%
\bibitem [{\citenamefont {Kantowski}\ \emph {et~al.}(2012)\citenamefont {Kantowski}, \citenamefont {Chen},\ and\ \citenamefont {Dai}}]{Kant2}%
  \BibitemOpen
  \bibfield  {author} {\bibinfo {author} {\bibfnamefont {R.}~\bibnamefont {Kantowski}}, \bibinfo {author} {\bibfnamefont {B.}~\bibnamefont {Chen}},\ and\ \bibinfo {author} {\bibfnamefont {X.}~\bibnamefont {Dai}},\ }\bibfield  {title} {\bibinfo {title} {Image properties of embedded lenses},\ }\href {https://doi.org/10.1103/PhysRevD.86.043009} {\bibfield  {journal} {\bibinfo  {journal} {Phys. Rev. D}\ }\textbf {\bibinfo {volume} {86}},\ \bibinfo {pages} {083009} (\bibinfo {year} {2012})}\BibitemShut {NoStop}%
\bibitem [{\citenamefont {Arakida}\ and\ \citenamefont {Kasai}(2012)}]{Arakida}%
  \BibitemOpen
  \bibfield  {author} {\bibinfo {author} {\bibfnamefont {H.}~\bibnamefont {Arakida}}\ and\ \bibinfo {author} {\bibfnamefont {M.}~\bibnamefont {Kasai}},\ }\bibfield  {title} {\bibinfo {title} {Effect of the cosmological constant on the bending of light and the cosmological lens equation},\ }\href {https://doi.org/10.1103/PhysRevD.85.023006} {\bibfield  {journal} {\bibinfo  {journal} {Phys. Rev. D}\ }\textbf {\bibinfo {volume} {85}},\ \bibinfo {pages} {023006} (\bibinfo {year} {2012})}\BibitemShut {NoStop}%
\bibitem [{\citenamefont {Kantowski}\ \emph {et~al.}(2013)\citenamefont {Kantowski}, \citenamefont {Chen},\ and\ \citenamefont {Dai}}]{Kant3}%
  \BibitemOpen
  \bibfield  {author} {\bibinfo {author} {\bibfnamefont {R.}~\bibnamefont {Kantowski}}, \bibinfo {author} {\bibfnamefont {B.}~\bibnamefont {Chen}},\ and\ \bibinfo {author} {\bibfnamefont {X.}~\bibnamefont {Dai}},\ }\bibfield  {title} {\bibinfo {title} {Fermat’s least-time principle and the embedded transparent lens},\ }\href {https://doi.org/10.1103/PhysRevD.88.083001} {\bibfield  {journal} {\bibinfo  {journal} {Phys. Rev. D}\ }\textbf {\bibinfo {volume} {86}},\ \bibinfo {pages} {083001} (\bibinfo {year} {2013})}\BibitemShut {NoStop}%
\bibitem [{\citenamefont {Sultana}(2013)}]{Sultana1}%
  \BibitemOpen
  \bibfield  {author} {\bibinfo {author} {\bibfnamefont {J.}~\bibnamefont {Sultana}},\ }\bibfield  {title} {\bibinfo {title} {Contribution of the cosmological constant to the bending of light in {Kerr–de Sitter} spacetime},\ }\href {https://doi.org/10.1103/PhysRevD.88.042003} {\bibfield  {journal} {\bibinfo  {journal} {Phys. Rev. D}\ }\textbf {\bibinfo {volume} {88}},\ \bibinfo {pages} {042003} (\bibinfo {year} {2013})}\BibitemShut {NoStop}%
\bibitem [{\citenamefont {Heydari-Fard}\ \emph {et~al.}(2021)\citenamefont {Heydari-Fard}, \citenamefont {Heydari-Fard},\ and\ \citenamefont {Sepang}}]{Heydari}%
  \BibitemOpen
  \bibfield  {author} {\bibinfo {author} {\bibfnamefont {M.}~\bibnamefont {Heydari-Fard}}, \bibinfo {author} {\bibfnamefont {M.}~\bibnamefont {Heydari-Fard}},\ and\ \bibinfo {author} {\bibfnamefont {H.~R.}\ \bibnamefont {Sepang}},\ }\bibfield  {title} {\bibinfo {title} {Bending of light in novel {4D} {Gauss-Bonnet-de Sitter} black holes by the {Rindler-Ishak} method},\ }\href {https://doi.org/10.1209/0295-5075/133/50006} {\bibfield  {journal} {\bibinfo  {journal} {Europhys. Lett.}\ }\textbf {\bibinfo {volume} {133}},\ \bibinfo {pages} {50006} (\bibinfo {year} {2021})}\BibitemShut {NoStop}%
\bibitem [{\citenamefont {Hu}\ \emph {et~al.}(2022)\citenamefont {Hu}, \citenamefont {Heavens},\ and\ \citenamefont {Bacon}}]{Hu}%
  \BibitemOpen
  \bibfield  {author} {\bibinfo {author} {\bibfnamefont {L.}~\bibnamefont {Hu}}, \bibinfo {author} {\bibfnamefont {A.}~\bibnamefont {Heavens}},\ and\ \bibinfo {author} {\bibfnamefont {D.}~\bibnamefont {Bacon}},\ }\bibfield  {title} {\bibinfo {title} {Light bending by the cosmological constant},\ }\href {https://doi.org/10.1088/1475-7516/2022/02/009} {\bibfield  {journal} {\bibinfo  {journal} {J. Cosmol. Astropart. Phys.}\ }\textbf {\bibinfo {volume} {02}}\bibinfo  {number} { (2022)},\ \bibinfo {pages} {009}}\BibitemShut {NoStop}%
\bibitem [{\citenamefont {Sultana}(2023)}]{Sultana2}%
  \BibitemOpen
\bibfield  {number} {  }\bibfield  {author} {\bibinfo {author} {\bibfnamefont {J.}~\bibnamefont {Sultana}},\ }\bibfield  {title} {\bibinfo {title} {Comment on “gravitational lensing in {Weyl} gravity”},\ }\href {https://doi.org/10.1103/PhysRevD.108.108501} {\bibfield  {journal} {\bibinfo  {journal} {Phys. Rev. D}\ }\textbf {\bibinfo {volume} {108}},\ \bibinfo {pages} {108501} (\bibinfo {year} {2023})}\BibitemShut {NoStop}%
\bibitem [{\citenamefont {Sultana}(2024)}]{Sultana3}%
  \BibitemOpen
  \bibfield  {author} {\bibinfo {author} {\bibfnamefont {J.}~\bibnamefont {Sultana}},\ }\bibfield  {title} {\bibinfo {title} {Gravitational light bending in {Weyl} gravity and {Schwarzschild–de Sitter} spacetime},\ }\href {https://doi.org/10.3390/sym16010101} {\bibfield  {journal} {\bibinfo  {journal} {Symmetry}\ }\textbf {\bibinfo {volume} {16}},\ \bibinfo {pages} {101} (\bibinfo {year} {2024})}\BibitemShut {NoStop}%
\bibitem [{\citenamefont {Lu}\ \emph {et~al.}(2025)\citenamefont {Lu}, \citenamefont {Pan}, \citenamefont {Lai},\ and\ \citenamefont {Wang}}]{Wang}%
  \BibitemOpen
  \bibfield  {author} {\bibinfo {author} {\bibfnamefont {Y.}~\bibnamefont {Lu}}, \bibinfo {author} {\bibfnamefont {X.-Y.}\ \bibnamefont {Pan}}, \bibinfo {author} {\bibfnamefont {M.-Y.}\ \bibnamefont {Lai}},\ and\ \bibinfo {author} {\bibfnamefont {Q.-H.}\ \bibnamefont {Wang}},\ }\href@noop {} {\bibinfo {title} {Finite-distance gravitational lensing of a global monopole in {Schwarzschild-de Sitter} spacetime}} (\bibinfo {year} {2025}),\ \Eprint {https://arxiv.org/abs/arXiv:2504.00777} {arXiv:2504.00777} \BibitemShut {NoStop}%
\bibitem [{\citenamefont {Schücker}(2009{\natexlab{b}})}]{Schu1}%
  \BibitemOpen
  \bibfield  {author} {\bibinfo {author} {\bibfnamefont {T.}~\bibnamefont {Schücker}},\ }\bibfield  {title} {\bibinfo {title} {Strong lensing in the {Einstein-Straus} solution},\ }\href {https://doi.org/10.1007/s10714-008-0731-4} {\bibfield  {journal} {\bibinfo  {journal} {Gen. Relativ. Gravit.}\ }\textbf {\bibinfo {volume} {41}},\ \bibinfo {pages} {1595} (\bibinfo {year} {2009}{\natexlab{b}})}\BibitemShut {NoStop}%
\bibitem [{\citenamefont {Boudjemaa}\ \emph {et~al.}(2011)\citenamefont {Boudjemaa}, \citenamefont {Guenouche},\ and\ \citenamefont {Zouzou}}]{Guen1}%
  \BibitemOpen
  \bibfield  {author} {\bibinfo {author} {\bibfnamefont {K.-E.}\ \bibnamefont {Boudjemaa}}, \bibinfo {author} {\bibfnamefont {M.}~\bibnamefont {Guenouche}},\ and\ \bibinfo {author} {\bibfnamefont {S.~R.}\ \bibnamefont {Zouzou}},\ }\bibfield  {title} {\bibinfo {title} {Time delay in the {Einstein-Straus} solution},\ }\href {https://doi.org/10.1007/s10714-011-1152-3} {\bibfield  {journal} {\bibinfo  {journal} {Gen. Relativ. Gravit.}\ }\textbf {\bibinfo {volume} {43}},\ \bibinfo {pages} {1707} (\bibinfo {year} {2011})}\BibitemShut {NoStop}%
\bibitem [{\citenamefont {Guenouche}\ and\ \citenamefont {Zouzou}(2018)}]{Guen2}%
  \BibitemOpen
  \bibfield  {author} {\bibinfo {author} {\bibfnamefont {M.}~\bibnamefont {Guenouche}}\ and\ \bibinfo {author} {\bibfnamefont {S.~R.}\ \bibnamefont {Zouzou}},\ }\bibfield  {title} {\bibinfo {title} {Deflection of light and time delay in closed {Einstein-Straus} solution},\ }\href {https://doi.org/10.1103/PhysRevD.98.123508} {\bibfield  {journal} {\bibinfo  {journal} {Phys. Rev. D}\ }\textbf {\bibinfo {volume} {98}},\ \bibinfo {pages} {123508} (\bibinfo {year} {2018})}\BibitemShut {NoStop}%
\bibitem [{\citenamefont {Schücker}\ and\ \citenamefont {Zaimen}(2008)}]{Schu2}%
  \BibitemOpen
  \bibfield  {author} {\bibinfo {author} {\bibfnamefont {T.}~\bibnamefont {Schücker}}\ and\ \bibinfo {author} {\bibfnamefont {N.}~\bibnamefont {Zaimen}},\ }\bibfield  {title} {\bibinfo {title} {Cosmological constant and time delay},\ }\href {https://doi.org/10.1051/0004-6361:200809449} {\bibfield  {journal} {\bibinfo  {journal} {Astron. Astrophys.}\ }\textbf {\bibinfo {volume} {484}},\ \bibinfo {pages} {103} (\bibinfo {year} {2008})}\BibitemShut {NoStop}%
\bibitem [{\citenamefont {Inada}\ \emph {et~al.}(2003)\citenamefont {Inada} \emph {et~al.}}]{Inada}%
  \BibitemOpen
  \bibfield  {author} {\bibinfo {author} {\bibfnamefont {N.}~\bibnamefont {Inada}} \emph {et~al.},\ }\bibfield  {title} {\bibinfo {title} {A gravitationally lensed quasar with quadruple images separated by 14.62 arcseconds},\ }\href {https://doi.org/10.1038/nature02153} {\bibfield  {journal} {\bibinfo  {journal} {Nature}\ }\textbf {\bibinfo {volume} {426}},\ \bibinfo {pages} {810} (\bibinfo {year} {2003})}\BibitemShut {NoStop}%
\bibitem [{\citenamefont {Oguri}\ \emph {et~al.}(2004)\citenamefont {Oguri} \emph {et~al.}}]{Oguri}%
  \BibitemOpen
  \bibfield  {author} {\bibinfo {author} {\bibfnamefont {M.}~\bibnamefont {Oguri}} \emph {et~al.},\ }\bibfield  {title} {\bibinfo {title} {Observations and theoretical implications of the large separation lensed quasar {SDSS J1004+4112}},\ }\href {https://doi.org/10.1086/382221} {\bibfield  {journal} {\bibinfo  {journal} {Astrophys. J.}\ }\textbf {\bibinfo {volume} {605}},\ \bibinfo {pages} {78} (\bibinfo {year} {2004})}\BibitemShut {NoStop}%
\bibitem [{\citenamefont {Fohlmeister}\ \emph {et~al.}(2008)\citenamefont {Fohlmeister}, \citenamefont {Kochanek}, \citenamefont {Falco}, \citenamefont {Morgan},\ and\ \citenamefont {Wambsganss}}]{Fohl}%
  \BibitemOpen
  \bibfield  {author} {\bibinfo {author} {\bibfnamefont {J.}~\bibnamefont {Fohlmeister}}, \bibinfo {author} {\bibfnamefont {C.~S.}\ \bibnamefont {Kochanek}}, \bibinfo {author} {\bibfnamefont {E.~E.}\ \bibnamefont {Falco}}, \bibinfo {author} {\bibfnamefont {C.~W.}\ \bibnamefont {Morgan}},\ and\ \bibinfo {author} {\bibfnamefont {J.}~\bibnamefont {Wambsganss}},\ }\bibfield  {title} {\bibinfo {title} {The rewards of patience: An 822 day time delay in the gravitational lens {SDSS J1004+4112}},\ }\href {https://doi.org/10.1086/528789} {\bibfield  {journal} {\bibinfo  {journal} {Astrophys. J.}\ }\textbf {\bibinfo {volume} {676}},\ \bibinfo {pages} {761} (\bibinfo {year} {2008})}\BibitemShut {NoStop}%
\bibitem [{\citenamefont {Kawano}\ and\ \citenamefont {Oguri}(2006)}]{Kawano}%
  \BibitemOpen
  \bibfield  {author} {\bibinfo {author} {\bibfnamefont {Y.}~\bibnamefont {Kawano}}\ and\ \bibinfo {author} {\bibfnamefont {M.}~\bibnamefont {Oguri}},\ }\bibfield  {title} {\bibinfo {title} {Time delays for the giant quadruple lensed {SDSS J1004+4112}: Prospects for determining the density profile of the lensing cluster},\ }\href {https://doi.org/10.1093/pasj/58.2.271} {\bibfield  {journal} {\bibinfo  {journal} {Publ. Astron. Soc. Jpn.}\ }\textbf {\bibinfo {volume} {58}},\ \bibinfo {pages} {271} (\bibinfo {year} {2006})}\BibitemShut {NoStop}%
\bibitem [{\citenamefont {Schücker}(2010{\natexlab{b}})}]{Schu3}%
  \BibitemOpen
  \bibfield  {author} {\bibinfo {author} {\bibfnamefont {T.}~\bibnamefont {Schücker}},\ }\bibfield  {title} {\bibinfo {title} {Lensing in an interior {Kottler} solution},\ }\href {https://doi.org/10.1007/s10714-010-0978-4} {\bibfield  {journal} {\bibinfo  {journal} {Gen. Relativ. Gravit.}\ }\textbf {\bibinfo {volume} {42}},\ \bibinfo {pages} {1991} (\bibinfo {year} {2010}{\natexlab{b}})}\BibitemShut {NoStop}%
\bibitem [{\citenamefont {Guenouche}(2025{\natexlab{a}})}]{Guen4}%
  \BibitemOpen
  \bibfield  {author} {\bibinfo {author} {\bibfnamefont {M.}~\bibnamefont {Guenouche}},\ }\href@noop {} {\bibinfo {title} {Lensing in matched exterior and interior {Kottler} solutions}} (\bibinfo {year} {2025}{\natexlab{a}}),\ \Eprint {https://arxiv.org/abs/arXiv:2509.00630} {arXiv:2509.00630} \BibitemShut {NoStop}%
\bibitem [{\citenamefont {Guenouche}(2025{\natexlab{b}})}]{Guen5}%
  \BibitemOpen
  \bibfield  {author} {\bibinfo {author} {\bibfnamefont {M.}~\bibnamefont {Guenouche}},\ }\href@noop {} {\bibinfo {title} {Time delay in matched exterior and interior {Kottler} solutions}} (\bibinfo {year} {2025}{\natexlab{b}}),\ \Eprint {https://arxiv.org/abs/arXiv:2509.11049} {arXiv:2509.11049} \BibitemShut {NoStop}%
\bibitem [{\citenamefont {Virbhadra}\ and\ \citenamefont {Ellis}(2000)}]{Virbhadra2000}%
  \BibitemOpen
  \bibfield  {author} {\bibinfo {author} {\bibfnamefont {K.~S.}\ \bibnamefont {Virbhadra}}\ and\ \bibinfo {author} {\bibfnamefont {G.~F.~R.}\ \bibnamefont {Ellis}},\ }\bibfield  {title} {\bibinfo {title} {Schwarzschild black hole lensing},\ }\href {https://doi.org/10.1103/PhysRevD.62.084003} {\bibfield  {journal} {\bibinfo  {journal} {Phys. Rev. D}\ }\textbf {\bibinfo {volume} {62}},\ \bibinfo {pages} {084003} (\bibinfo {year} {2000})}\BibitemShut {NoStop}%
\bibitem [{\citenamefont {Virbhadra}\ and\ \citenamefont {Keeton}(2008)}]{Virbhadra2002}%
  \BibitemOpen
  \bibfield  {author} {\bibinfo {author} {\bibfnamefont {K.~S.}\ \bibnamefont {Virbhadra}}\ and\ \bibinfo {author} {\bibfnamefont {C.~R.}\ \bibnamefont {Keeton}},\ }\bibfield  {title} {\bibinfo {title} {Time delay and magnification centroid due to gravitational lensing by black holes and naked singularities},\ }\href {https://doi.org/10.1103/PhysRevD.77.124014} {\bibfield  {journal} {\bibinfo  {journal} {Phys. Rev. D}\ }\textbf {\bibinfo {volume} {77}},\ \bibinfo {pages} {124014} (\bibinfo {year} {2008})}\BibitemShut {NoStop}%
\bibitem [{\citenamefont {Virbhadra}(2009)}]{VirbhadraEllis2009}%
  \BibitemOpen
  \bibfield  {author} {\bibinfo {author} {\bibfnamefont {K.~S.}\ \bibnamefont {Virbhadra}},\ }\bibfield  {title} {\bibinfo {title} {Relativistic images of {Schwarzschild} black hole lensing},\ }\href {https://doi.org/10.1103/PhysRevD.79.083004} {\bibfield  {journal} {\bibinfo  {journal} {Phys. Rev. D}\ }\textbf {\bibinfo {volume} {79}},\ \bibinfo {pages} {083004} (\bibinfo {year} {2009})}\BibitemShut {NoStop}%
\bibitem [{\citenamefont {Battista}\ \emph {et~al.}(2017)\citenamefont {Battista} \emph {et~al.}}]{Battista2017}%
  \BibitemOpen
  \bibfield  {author} {\bibinfo {author} {\bibfnamefont {E.}~\bibnamefont {Battista}} \emph {et~al.},\ }\bibfield  {title} {\bibinfo {title} {Quantum time delay in the gravitational field of a rotating mass},\ }\href {https://doi.org/10.1088/1361-6382/aa7f11} {\bibfield  {journal} {\bibinfo  {journal} {Class. Quant. Grav.}\ }\textbf {\bibinfo {volume} {34}},\ \bibinfo {pages} {165008} (\bibinfo {year} {2017})}\BibitemShut {NoStop}%
\bibitem [{\citenamefont {Soares}\ \emph {et~al.}(2023{\natexlab{a}})\citenamefont {Soares}, \citenamefont {Pereira}, \citenamefont {Vitória},\ and\ \citenamefont {Rocha}}]{Soares2023a}%
  \BibitemOpen
  \bibfield  {author} {\bibinfo {author} {\bibfnamefont {A.~R.}\ \bibnamefont {Soares}}, \bibinfo {author} {\bibfnamefont {C.~F.~S.}\ \bibnamefont {Pereira}}, \bibinfo {author} {\bibfnamefont {R.~L.~L.}\ \bibnamefont {Vitória}},\ and\ \bibinfo {author} {\bibfnamefont {E.~M.}\ \bibnamefont {Rocha}},\ }\bibfield  {title} {\bibinfo {title} {Holonomy corrected {Schwarzschild} black hole lensing},\ }\href {https://doi.org/10.1103/PhysRevD.108.124024} {\bibfield  {journal} {\bibinfo  {journal} {Phys. Rev. D}\ }\textbf {\bibinfo {volume} {108}},\ \bibinfo {pages} {124024} (\bibinfo {year} {2023}{\natexlab{a}})}\BibitemShut {NoStop}%
\bibitem [{\citenamefont {Soares}\ \emph {et~al.}(2024)\citenamefont {Soares}, \citenamefont {Vitória},\ and\ \citenamefont {Pereira}}]{Soares2023b}%
  \BibitemOpen
  \bibfield  {author} {\bibinfo {author} {\bibfnamefont {A.~R.}\ \bibnamefont {Soares}}, \bibinfo {author} {\bibfnamefont {R.~L.~L.}\ \bibnamefont {Vitória}},\ and\ \bibinfo {author} {\bibfnamefont {C.~F.~S.}\ \bibnamefont {Pereira}},\ }\bibfield  {title} {\bibinfo {title} {Topologically charged holonomy corrected {Schwarzschild} black hole lensing},\ }\href {https://doi.org/10.1103/PhysRevD.110.084004} {\bibfield  {journal} {\bibinfo  {journal} {Phys. Rev. D}\ }\textbf {\bibinfo {volume} {110}},\ \bibinfo {pages} {084004} (\bibinfo {year} {2024})}\BibitemShut {NoStop}%
\bibitem [{\citenamefont {Soares}\ \emph {et~al.}(2023{\natexlab{b}})\citenamefont {Soares}, \citenamefont {Vitória},\ and\ \citenamefont {Pereira}}]{Soares2023c}%
  \BibitemOpen
  \bibfield  {author} {\bibinfo {author} {\bibfnamefont {A.~R.}\ \bibnamefont {Soares}}, \bibinfo {author} {\bibfnamefont {R.~L.~L.}\ \bibnamefont {Vitória}},\ and\ \bibinfo {author} {\bibfnamefont {C.~F.~S.}\ \bibnamefont {Pereira}},\ }\bibfield  {title} {\bibinfo {title} {Gravitational lensing in a topologically charged {Eddington}-inspired {Born–Infeld} spacetime},\ }\href {https://doi.org/10.1140/epjc/s10052-023-12071-z} {\bibfield  {journal} {\bibinfo  {journal} {Eur. Phys. J. C}\ }\textbf {\bibinfo {volume} {83}},\ \bibinfo {pages} {903} (\bibinfo {year} {2023}{\natexlab{b}})}\BibitemShut {NoStop}%
\bibitem [{\citenamefont {Soares}\ \emph {et~al.}(2025)\citenamefont {Soares} \emph {et~al.}}]{Soares2023d}%
  \BibitemOpen
  \bibfield  {author} {\bibinfo {author} {\bibfnamefont {A.~R.}\ \bibnamefont {Soares}} \emph {et~al.},\ }\bibfield  {title} {\bibinfo {title} {Light deflection and gravitational lensing effects inspired by loop quantum gravity},\ }\href {https://doi.org/10.1088/1475-7516/2025/06/034} {\bibfield  {journal} {\bibinfo  {journal} {J. Cosmol. Astropart. Phys.}\ }\textbf {\bibinfo {volume} {06}}\bibinfo  {number} { (2025)},\ \bibinfo {pages} {034}}\BibitemShut {NoStop}%
\end{thebibliography}%

\end{document}